# Algorithm for solving a pump-probe model for an arbitrary number of energy levels

Zifan Zhou[1,*], Yael Sternfeld[2,*], Jacob Scheuer[3] and M. S. Shahriar[1,4]

[1] *Northwestern University, Department of Electrical and Computer Engineering, Evanston, IL, 60208, USA.*
[2] *Tel Aviv University, Department of Physics and Astronomy, Ramat-Aviv, Tel Aviv 69978, Israel.*
[3] *Tel Aviv University, School of Electrical Engineering, Faculty of Engineering, Ramat-Aviv, Tel Aviv 69978, Israel.*
[4] *Northwestern University, Department of Physics and Astronomy, Evanston, IL, 60208, USA.*
* *The authors contributed equally to this work.*

**Abstract**: We describe a generalized algorithm for evaluating the steady-state solution of the density matrix equation of motion, for the pump-probe scheme, when two fields oscillating at different frequencies couple the same set of atomic transitions involving an arbitrary number of energy levels, to an arbitrary order of the harmonics of the pump-probe frequency difference. We developed a numerical approach and a symbolic approach for this algorithm. We have verified that both approaches yield the same result for all cases studied, but require different computation time. The results are further validated by comparing them with the analytical solution of a two-level system to first order. We have also used both models to produce results up to the third order in the pump-probe frequency difference, for two-, three- and four-level systems. In addition, we have used this model to determine accurately, for the first time, the gain profile for a self-pumped Raman laser, for a system involving 16 Zeeman sublevels in the D1 manifold of $^{87}$Rb atoms. We have also used this model to determine the behavior of a single-pumped superluminal laser. In many situations involving the applications of multiple laser fields to atoms with many energy levels, one often makes the approximation that each field couples only one transition, because of the difficulty encountered in accounting for the effect of another field coupling the same transition but with a large detuning. The use of the algorithm presented here would eliminate the need for making such approximations, thus improving the accuracy of numerical calculations for such schemes.

## 1. Introduction

In the semi-classical model for atom-laser interactions, the atoms are treated quantum mechanically while the light fields are treated classically. In the simplest case, only one monochromatic field is used to excite a single optical transition. Under this condition, the rotating wave approximation (RWA) and the rotating wave transformation (RWT) can be applied to derive a time-independent Hamiltonian, which significantly simplifies the procedure of solving the density matrix equation of motion. However, in cases where more than a single frequency drives the same transition, the RWT cannot fully eliminate the time dependent terms in the Hamiltonian. Thus, in such cases the Hamiltonian contains time-oscillating terms [1,2,3,4,5,6,7]. The quasi-steady-state solution of the density matrix driven by such a Hamiltonian includes essentially infinite number of harmonics of the frequency difference between the two driving fields [8,9]. In the general case, the equations are solved by keeping harmonic terms up to a value that is sufficiently large so that adding one more term produces insignificant changes in the solutions. However, this approach becomes exceedingly complex when many levels are involved.

Consider, for example, the case using $^{87}$Rb atoms where one laser frequency is tuned close to resonance with the transition between the $5S_{1/2}$, F=1 state and the $5P_{1/2}$, F=1 excited state, and another laser frequency is tuned close to resonance with the transition between the $5S_{1/2}$, F=2 state and the same excited state.  Due to the presence of the Zeeman sublevels within each of these hyperfine states, each field will cause coupling along both transitions, for any combination of polarizations of these fields.  In such a situation, it is customary to make the simplifying assumption that each of these frequencies act only along the transition that is close to resonance.  However, this approximation limits the precision of the model, especially when the Rabi frequencies are not very small compared to the ground-state hyperfine splitting.  Furthermore, this approximation breaks down when the detuning of the field becomes comparable to the ground-state hyperfine splitting.  In such scenarios, it is necessary to account for the fact that each transition is being excited by fields at two different frequencies.  The resulting analysis can become prohibitively difficult when a large number of energy levels (e.g., the Zeeman sublevels in the example mentioned above) are involved.  As a result, it is customary for scientists to continue to make the above-mentioned simplifying assumption.  It should be noted that while we have illustrated the issue using the case of a system involving three hyperfine states, it is relevant in virtually all systems subjected to excitation by more than one laser frequency.

In this paper, we present an algorithm that can be used to carry out the proper analysis in such scenarios without using the above-mentioned approximation, for a system involving an arbitrary number of energy levels, while keeping terms up to an arbitrary order of the difference frequency.  We developed numerical and symbolic approaches for this algorithm. These approaches are built upon the framework of an approach we had developed earlier [10] for automated generation of the density matrix equations of motion for a system with an arbitrary number of energy levels, limited to conditions where each frequency is assumed to excite only one transition. We have verified that both the numerical and the symbolic approaches yield the same result for all cases studied, but generally require different computation time. The results are further validated by comparing them with the analytical solution of a two-level system to first order.  We have also used both models to analyze a multi-frequency driven multi-level system up to the third order in the pump-probe frequency difference, for two-, three- and four-level systems. Finally, we have used this model to determine accurately, for the first time, the gain profile for a self-pumped Raman laser [11,12], for a system involving 16 Zeeman sublevels in the D1 manifold of $^{87}$Rb atoms.

The complexity resulting from the excitation of the same hyperfine transition using two frequencies is of immediate and practical relevance in our recent investigation into simplification of schemes for realizing a superluminal laser [13]. It has been shown [14,15,16,17,18,19,20] that the output frequency of a superluminal laser is extremely sensitive to rotation and cavity length perturbations, which can be employed for precision metrology. The realization of a superluminal laser requires the gain spectrum to have a narrow dip at the center of a broad gain profile. Over the years, to produce such a gain profile, various approaches have been developed and investigated by different groups [9,21,22,23,24,25,26,27,28,29]. Most recently, we have identified a very simple approach for realizing a superluminal ring laser using a single isotope of Rb and a single pump laser, by producing electro-magnetically induced transparency (EIT) in the self-pumped Raman gain scheme. In this approach, which is summarized in Figure 1 of Ref. [13], the Raman pump produces the ground state population inversion for Raman gain. In addition, it produces a dip in the gain profile via Autler-Townes splitting of the transition that is resonant with the Raman pump. In Ref [13], we described experimental realization of the superluminal laser gain profiles in both D1 and D2 manifolds. We also presented a theoretical model for describing this process. However, this model was an approximate one, relying heavily on several fitting parameters. In order to develop a proper model, it is necessary to address the complexity due to the fact that, under this scheme, the same transition is coupled by both the Raman pump field and the Raman probe field, which are at different frequencies. Furthermore, a proper model must take into account all hyperfine levels as well as Zeeman sublevels. The 4-level case presented in this paper shows how to apply these algorithms to predict accurately the gain profile for such a single-pumped superluminal laser.

Another important application of the technique developed in this paper would be to determine accurately the process of radiation trapping in a magneto-optic trap (MOT). Recalling briefly, this is a process whereby an atom experiences an attraction to or repulsion from a neighboring atom, depending on whether it amplifies or absorbs light emitted by the neighbor. In a MOT, every atom is irradiated by the fluorescence emitted by all the neighboring atoms. The fluorescence spectrum of each atom can be calculated with relative ease by applying the quantum regression theorem [30,31,32,33], even when taking into account the multi-level structure of the atoms, since the excitation field is monochromatic. However, accurate determination of the absorption-or-gain spectrum of the atoms is difficult because of the need to take into account the multi-level structure, especially when the fluorescence produced by the neighboring atoms is strong enough to require keeping track of many orders of the pump-probe beat frequency. The method described in this paper can be applied to determine the absorption-or-gain spectrum of trapped atoms very accurately, thereby determining the nature of the radiation trapping force with high precision. This information in turn can possibly be used to optimize density of atoms in a MOT, as well as tailor the three-dimensional distribution of atoms.

The rest of the paper is organized as follows. In Section 2, we describe the numerical approach for solving the steady state density matrix using a 2-level system as an example. In Section 3, the symbolic approach is discussed. In Section 4, we show the results for the 3-level system generated by the two approaches. In Section 5, we show the results for the 4-level system, which describes the basic behavior of the gain profile in a single-pumped superluminal laser. In Section 6, we generate the gain profile for a self-pumped Raman laser, for a system involving 16 Zeeman sublevels. Discussions and conclusions are presented in Section 7. In Appendix A, we describe the generalization of the numerical approach for a system with an arbitrary number of energy levels and keeping arbitrary orders of time oscillating terms. In Appendix B, we show the MATLAB code for implementing the numerical algorithm, and in Appendix C, we show the MATLAB code for implementing the symbolic algorithm.

## 2. Numerical approach

For illustrating the algorithm under the simplest possible condition, we consider first a two-level system that is identical to the one presented in Ref. [8], which is driven by a strong pump and a weak probe. It also allows for optical pumping from the ground to the excited state that is sufficiently strong to produce population inversion between these states. The schematic of the energy levels and optical fields is shown in Figure 1. We start with the Liouville equation:

$$\frac{\partial \tilde{\rho}}{\partial t} = -\frac{i}{\hbar}\left[\tilde{\tilde{H}}\tilde{\rho} - \tilde{\rho}\tilde{\tilde{H}}^{\dagger}\right] + \tilde{\rho}_s, \tag{1}$$

$$\tilde{\tilde{H}} = \frac{\hbar}{2}\begin{bmatrix} -i\Gamma_{op} & \Omega_p + \Omega_s e^{i\delta t} \\ \Omega_p + \Omega_s e^{-i\delta t} & -2\Delta - i\Gamma \end{bmatrix}, \tag{2}$$

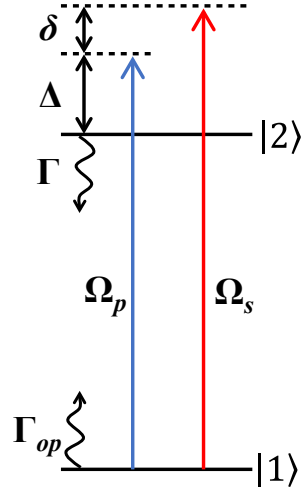

Figure 1. Schematic of the energy levels and the optical fields in a two-level system driven by two different frequencies.

$$\tilde{\rho}_s = \begin{bmatrix} \Gamma \tilde{\rho}_{22} & 0 \\ 0 & \Gamma_{op} \tilde{\rho}_{11} \end{bmatrix}. \tag{3}$$

Here, $\tilde{H}$ is the Hamiltonian after the rotating wave approximation (RWA) and rotating wave transformation (RWT), $\tilde{\rho}$ is the density matrix after the RWA and the RWT, $\tilde{\rho}_s$ represents the source terms that account for the influx of the atoms due to decay, $\Gamma_{op}$ is the effective incoherent excitation rate induced by a mechanism such as optical pumping, $\Omega_p$ and $\Omega_s$ are the Rabi frequencies of the pump field and the probe field, respectively, $\Delta$ is the detuning of the pump field with a respect to the resonance of the $|1\rangle \leftrightarrow |2\rangle$ transition, $\delta$ is the frequency difference between the probe field and the pump field, and $\Gamma$ is the atomic decay from level $|2\rangle$. It should be noted that in this formulation the Hamiltonian includes the decay terms; as such, it is non-Hermitian. Due to the presence of two fields with different frequencies applied on a single transition, the RWT cannot eliminate the time varying terms in the Hamiltonian completely, as can be seen in Eq. (2). As a result, the density matrix cannot have a fully steady-state solution. For most practical situations, what is of interest is the *pseudo steady-state solution*, under which each element of the density matrix will be a sum of many terms, including a stationary term and terms oscillating at all the harmonics (positive and negative) of $\delta$. In other words, each element of the density matrix will be periodic in time, with a period given by the inverse of the frequency difference ($\delta$) between the pump and the probe. To start with, we only consider the constant terms and the first order (positive and negative) terms. This approximation is adequate to describe conditions where the probe is infinitesimally weak. In what follows, we will denote this as the weak-probe case. Later on, we will consider the more general case where harmonics up to an arbitrary order are taken into account. Under this approximation, we can write:

$$\tilde{\rho} = \tilde{\rho}^0 + \tilde{\rho}^1 e^{i\delta t} + \tilde{\rho}^{-1} e^{-i\delta t}, \tag{4}$$

$$\tilde{\rho}_s = \tilde{\rho}_s^0 + \tilde{\rho}_s^1 e^{i\delta t} + \tilde{\rho}_s^{-1} e^{-i\delta t}. \tag{5}$$

(Before proceeding, we want to point out that in this paper superscripts on variables do not represent exponents.) The Liouville equation in pseudo steady state can be expressed as:

$$\frac{\partial \tilde{\rho}}{\partial t} = 0 + i\delta \tilde{\rho}^1 e^{i\delta t} - i\delta \tilde{\rho}^{-1} e^{-i\delta t} = -\frac{i}{\hbar}\left[\tilde{H}\tilde{\rho} - \tilde{\rho}\tilde{H}^\dagger\right] + \tilde{\rho}_s. \tag{6}$$

We will assume that the system is closed, so that the following constraints must be satisfied:

$$\tilde{\rho}_{11}^0 + \tilde{\rho}_{22}^0 = 1, \quad \tilde{\rho}_{11}^1 + \tilde{\rho}_{22}^1 = 0, \quad \tilde{\rho}_{11}^{-1} + \tilde{\rho}_{22}^{-1} = 0. \tag{7}$$

Similar to the N-level algorithm presented in [10], we need to convert the Liouville equation to a set of linear equations. For a two-level system, the number of linear equations for the weak-probe case is 12. For an arbitrary number of levels $N$, the corresponding number of linear equations is $3 \cdot N^2$. We now describe the algorithmic steps used to evaluate the coefficients in a 12×12 matrix $M$, which is time independent, satisfying the following equation:

$$MA = B, \tag{8}$$

where $B$ is the 12×1 null vector, and $A$ is the vectorized density matrix with all the components for each density matrix element. The order of the elements can be chosen arbitrarily. We choose to use the following ordering:

$$A \equiv \left[\tilde{\rho}_{11}^0, \tilde{\rho}_{11}^1, \tilde{\rho}_{11}^{-1}, \tilde{\rho}_{12}^0, \tilde{\rho}_{12}^1, \tilde{\rho}_{12}^{-1}, \tilde{\rho}_{21}^0, \tilde{\rho}_{21}^1, \tilde{\rho}_{21}^{-1}, \tilde{\rho}_{22}^0, \tilde{\rho}_{22}^1, \tilde{\rho}_{22}^{-1}\right]^T. \tag{9}$$

In principle, this ordering can be arbitrary. However, when the process of generalizing the algorithm to an arbitrary number of energy levels for an arbitrary number of orders, it is necessary to use a specific rule for creating the order. Here, we have used the rule explained below: using the example of a 2-level system:

The general form of the density matrix for the 2-level system is:

$$\tilde{\rho} = \begin{bmatrix} \left(\tilde{\rho}_{11}^0 + \tilde{\rho}_{11}^1 e^{i\delta t} + \tilde{\rho}_{11}^{-1} e^{-i\delta t} + \cdots\right) & \left(\tilde{\rho}_{12}^0 + \tilde{\rho}_{12}^1 e^{i\delta t} + \tilde{\rho}_{12}^{-1} e^{-i\delta t} + \cdots\right) \\ \left(\tilde{\rho}_{21}^0 + \tilde{\rho}_{21}^1 e^{i\delta t} + \tilde{\rho}_{21}^{-1} e^{-i\delta t} + \cdots\right) & \left(\tilde{\rho}_{22}^0 + \tilde{\rho}_{22}^1 e^{i\delta t} + \tilde{\rho}_{22}^{-1} e^{-i\delta t} + \cdots\right) \end{bmatrix}. \tag{10}$$

For each element, we have multiple harmonics, up to the order we need to consider, denoted as $K$. We arrange them from lower order to higher order, and for the same order, the negative harmonic follows the positive harmonic. We start with the first row and first column in the density matrix, and the coefficients multiplying the time oscillating factors in the order expressed in Eq. (10) form the first $(2K+1)$ elements in the $A$ vector. Then we move across the columns in the first row and fill the coefficients in the $A$ vector. After the first row is finished, we consider the second row and follow the same process until the entire density matrix is vectorized. For $N$ energy levels with keeping up to $K$ order of harmonics, the $A$ vector would be in the form of:

$$A \equiv \left[\tilde{\rho}_{11}^0, \tilde{\rho}_{11}^1, \tilde{\rho}_{11}^{-1}, \tilde{\rho}_{11}^2, \tilde{\rho}_{11}^{-2}, ..., \tilde{\rho}_{11}^K, \tilde{\rho}_{11}^{-K}, \tilde{\rho}_{12}^0, \tilde{\rho}_{12}^1, \tilde{\rho}_{12}^{-1}, ...., \tilde{\rho}_{NN}^K, \tilde{\rho}_{NN}^{-K}\right]^T. \tag{11}$$

It should be noted that the determinant of the matrix $M$ in Eq. (8) is zero, which is due to the fact that the 12 equations are not linearly independent, resulting from the closed-system constraints expressed in Eq. (7). These constraints can be used to eliminate 3 of the 12 equations, leading to a new equation of the form:

$$M'A' = B' \tag{12}$$

where $A'$ is a 9×1 column vector containing only 9 of the elements of the vector $A$, and $B'$ is a non-zero vector. Inversion of this equation, along with the constraints in Eq. (7), would give us all the elements of the vector $A$. It should also be noted that the time-independence of Eq. (8)

results from the fact that the time oscillating factors can be canceled out after rearranging the set of 12 equations derived from Eq. (6). The details of this process are described next.

First, we need to separate the terms in Eq. (6) that are multiplied by $e^{\pm i\delta t}$. For simplicity in notations, we introduce the following definitions:

$$\tilde{\tilde{H}} = H^0 + H^1 e^{i\delta t} + H^{-1} e^{-i\delta t}, \tag{13}$$

$$H^0 \equiv \frac{\hbar}{2}\begin{bmatrix} -i\Gamma_{op} & \Omega_p \\ \Omega_p & -\Delta - i\Gamma \end{bmatrix}, \tag{14}$$

$$H^1 \equiv \frac{\hbar}{2}\begin{bmatrix} 0 & \Omega_s \\ 0 & 0 \end{bmatrix}, \tag{15}$$

$$H^{-1} \equiv \frac{\hbar}{2}\begin{bmatrix} 0 & 0 \\ \Omega_s & 0 \end{bmatrix}. \tag{16}$$

Using these notations, the first term on the right-hand side of Eq. (6), which will be called the pseudo-commutator in the rest of the paper, can be expressed as:

$$\tilde{\tilde{H}}\tilde{\rho} - \tilde{\rho}\tilde{\tilde{H}}^\dagger = \left(H^0 + H^1 e^{i\delta t} + H^{-1} e^{-i\delta t}\right)\left(\tilde{\rho}^0 + \tilde{\rho}^1 e^{i\delta t} + \tilde{\rho}^{-1} e^{-i\delta t}\right) \\ - \left(\tilde{\rho}^0 + \tilde{\rho}^1 e^{i\delta t} + \tilde{\rho}^{-1} e^{-i\delta t}\right)\left(H^0 + H^1 e^{i\delta t} + H^{-1} e^{-i\delta t}\right)^\dagger. \tag{17}$$

Noting that $\left(H^1\right)^\dagger = H^{-1}$, we can write:

$$\left(H^0 + H^1 e^{i\delta t} + H^{-1} e^{-i\delta t}\right)^\dagger = \left(H^0\right)^\dagger + H^1 e^{i\delta t} + H^{-1} e^{-i\delta t}. \tag{18}$$

The pseudo-commutator of Eq. (17) can then be written as:

$$\tilde{\tilde{H}}\tilde{\rho} - \tilde{\rho}\tilde{\tilde{H}}^\dagger = \left(H^0 + H^1 e^{i\delta t} + H^{-1} e^{-i\delta t}\right)\left(\tilde{\rho}^0 + \tilde{\rho}^1 e^{i\delta t} + \tilde{\rho}^{-1} e^{-i\delta t}\right) \\ - \left(\tilde{\rho}^0 + \tilde{\rho}^1 e^{i\delta t} + \tilde{\rho}^{-1} e^{-i\delta t}\right)\left[\left(H^0\right)^\dagger + H^1 e^{i\delta t} + H^{-1} e^{-i\delta t}\right]. \tag{19}$$

In evaluating this pseudo-commutator, we will ignore higher order terms (such as those varying as $e^{\pm i 2\delta t}$). We group the remaining terms into the zeroth order, positive first order and negative first order as follows:

$$\tilde{\tilde{H}}\tilde{\rho} - \tilde{\rho}\tilde{\tilde{H}}^\dagger \approx U^0 + U^1 e^{i\delta t} + U^{-1} e^{-i\delta t}, \tag{20}$$

where

$$U^0 = \left[H^0 \tilde{\rho}^0 - \tilde{\rho}^0 \left(H^0\right)^\dagger\right] + \left(H^{-1}\tilde{\rho}^1 - \tilde{\rho}^1 H^{-1}\right) + \left(H^1 \tilde{\rho}^{-1} - \tilde{\rho}^{-1} H^1\right), \tag{21}$$

$$U^1 = \left(H^1 \tilde{\rho}^0 - \tilde{\rho}^0 H^1\right) + \left[H^0 \tilde{\rho}^1 - \tilde{\rho}^1 \left(H^0\right)^\dagger\right], \tag{22}$$

$$U^{-1} = \left(H^{-1}\tilde{\rho}^0 - \tilde{\rho}^0 H^{-1}\right) + \left[H^0 \tilde{\rho}^{-1} - \tilde{\rho}^{-1}\left(H^0\right)^\dagger\right]. \tag{23}$$

Equation (6) can now be rewritten as:

$$0 + i\delta\tilde{\rho}^1 e^{i\delta t} - i\delta\tilde{\rho}^{-1} e^{-i\delta t} = -\frac{i}{\hbar}\left(U^0 + U^1 e^{i\delta t} + U^{-1} e^{-i\delta t}\right) + \left(\tilde{\rho}_s^0 + \tilde{\rho}_s^1 e^{i\delta t} + \tilde{\rho}_s^{-1} e^{-i\delta t}\right). \tag{24}$$

As can be seen, we can group the terms that are multiplied by $e^{\pm i\delta t}$ in three equations:

$$G^0 \equiv -\frac{i}{\hbar}U^0 + \tilde{\rho}_s^0 = 0, \tag{25}$$

$$G^1 \equiv -\frac{i}{\hbar}U^1 + \tilde{\rho}_s^1 - i\delta\tilde{\rho}^1 = 0, \tag{26}$$

$$G^{-1} \equiv -\frac{i}{\hbar}U^{-1} + \tilde{\rho}_s^{-1} + i\delta\tilde{\rho}^{-1} = 0. \tag{27}$$

Each of Eq. (25) to Eq. (27) contains four linear equations. These can now be expressed as a set of 12 time-independent linear equations, in the form of Eq. (8). In what follows, we will denote these as equation $E_k$, with $k$ ranging from 1 through 12. Based on the ordering of the elements of $A$ vector shown earlier, the ordering for these linear equations would be as follows:

$$G_{11}^0 = 0 \Leftrightarrow E_1; \quad G_{12}^0 = 0 \Leftrightarrow E_4; \quad G_{21}^0 = 0 \Leftrightarrow E_7; \quad G_{22}^0 = 0 \Leftrightarrow E_{10}; \tag{28}$$

$$G_{11}^1 = 0 \Leftrightarrow E_2; \quad G_{12}^1 = 0 \Leftrightarrow E_5; \quad G_{21}^1 = 0 \Leftrightarrow E_8; \quad G_{22}^1 = 0 \Leftrightarrow E_{11}; \tag{29}$$

$$G_{11}^{-1} = 0 \Leftrightarrow E_3; \quad G_{12}^{-1} = 0 \Leftrightarrow E_6; \quad G_{21}^{-1} = 0 \Leftrightarrow E_9; \quad G_{22}^{-1} = 0 \Leftrightarrow E_{12}. \tag{30}$$

Next, we use a particular process, also used in Ref. [10], to determine the elements of the $M$ matrix. To illustrate the logic underlying this process, consider a generic situation where we have an equation of the form $V = ax + by + cz$ where the value of $V$ is a constant, and $x$, $y$ and $z$ are variables, and we want to determine the coefficients $a$, $b$ and $c$. It then follows that $a = V$ if we set the values of $x$ to unity and the values of $y$ and $z$ to zeroes, and so on. To use this process for finding the coefficients in $M$ matrix, we proceed as follows. For finding $M_{ij}$, we set the $j$-th element in the $A$ vector to unity, and the other elements to null values (it should be noted that the resulting $A$ vector does not correspond to a physically valid form of the density matrix; rather, this formulation is used as an algorithm step for extracting the coefficients in the $M$ matrix). The value of $M_{ij}$ is then given by the left-hand side (LHS) of equation $E_i$.

To illustrate this process, it is convenient to define first the following matrices:

$$Q^0 \equiv H^0 \begin{bmatrix} 1 & 0 \\ 0 & 0 \end{bmatrix} - \begin{bmatrix} 1 & 0 \\ 0 & 0 \end{bmatrix}(H^0)^\dagger; \tag{31}$$

$$Q^1 \equiv H^1 \begin{bmatrix} 1 & 0 \\ 0 & 0 \end{bmatrix} - \begin{bmatrix} 1 & 0 \\ 0 & 0 \end{bmatrix}H^1; \tag{32}$$

$$Q^{-1} \equiv H^{-1} \begin{bmatrix} 1 & 0 \\ 0 & 0 \end{bmatrix} - \begin{bmatrix} 1 & 0 \\ 0 & 0 \end{bmatrix}H^{-1}. \tag{33}$$

Let us consider the first linear equations (i.e., $E_1$) as an example of applying this process. Explicitly, the left-hand side (LHS) of $E_1$, denoted $E_{1(LHS)}$ as can be expressed as:

$$E_{1(LHS)} = G_{11}^0 = -\frac{i}{\hbar}U_{11}^0 + \tilde{\rho}_{s(11)}^0. \tag{34}$$

We set $\tilde{\rho}_{11}^0 = 1$ (corresponding to setting the first element of $A$ to unity) and all other elements in $A$ to zero. From Eq. (3), it then follows that $\tilde{\rho}_{s(11)}^0 = 0$. From Eq. (21) and Eq. (31), we see that $U_{11}^0 = Q_{11}^0$. Using these in Eq. (34), we find that $E_{1(LHS)} = (-i/\hbar)Q_{11}^0$, which in turn means that $M_{11} = E_{1(LHS)} = (-i/\hbar)Q_{11}^0$.

Similarly, the value of $M_{21}$ is given by the LHS of $E_2$, which can be expressed as:

$$E_{2(LHS)} = G_{11}^1 = -\frac{i}{\hbar}U_{11}^1 + \tilde{\rho}_{s(11)}^1 - i\delta\tilde{\rho}_{11}^1. \tag{35}$$

Since $\tilde{\rho}_{11}^0 = 1$ yields $\tilde{\rho}_{s(11)}^1 = 0$ and $\tilde{\rho}_{11}^1 = 0$, we have $E_{2(LHS)} = (-i/\hbar)Q_{11}^1$ from Eq. (22) and Eq. (32). As a result, we find that $M_{21} = E_{2(LHS)} = (-i/\hbar)Q_{11}^1$. For evaluating $M_{31}$, we calculate $G_{11}^{-1}$ with $\tilde{\rho}_{11}^0 = 1$, and we have $M_{31} = (-i/\hbar)Q_{11}^{-1}$.

For the rest of the elements in the first column in the $M$ matrix, we evaluate the LHS of $E_i$ ($i$ ranging from 4 to 12) with $\tilde{\rho}_{11}^0 = 1$. This procedure yields the following results:

$$M_{41} = E_{4(LHS)} = -\frac{i}{\hbar}Q_{12}^0; \quad M_{51} = E_{5(LHS)} = -\frac{i}{\hbar}Q_{12}^1; \quad M_{61} = E_{6(LHS)} = -\frac{i}{\hbar}Q_{12}^{-1}; \tag{36}$$

$$M_{71} = E_{7(LHS)} = -\frac{i}{\hbar}Q_{21}^0; \quad M_{81} = E_{8(LHS)} = -\frac{i}{\hbar}Q_{21}^1; \quad M_{91} = E_{9(LHS)} = -\frac{i}{\hbar}Q_{21}^{-1}; \tag{37}$$

$$M_{10,1} = E_{10(LHS)} = -\frac{i}{\hbar}Q_{22}^0 + \Gamma_{op}; \quad M_{11,1} = E_{11(LHS)} = -\frac{i}{\hbar}Q_{22}^1; \quad M_{12,1} = E_{12(LHS)} = -\frac{i}{\hbar}Q_{22}^{-1}. \tag{38}$$

Evaluation of the rest of the columns of the $M$ matrix can be carried out by following the same procedure. In this context, it should be noted that the three matrices defined in Eqs. (31), (32) and (33) are needed only for determining the elements of the first three columns of the $M$ matrix. A different set of three matrices, similar to those defined in Eqs. (31), (32) and (33), are needed for determining the elements of each of the three subsequent sets of three columns of the $M$ matrix.

Once all the elements of the $M$ matrix are determined, the next step makes use of the closed system constraint to remove the redundant equations within Eq. (8). By using Eq. (7), we find that the elements in the first three columns of the $M'$ matrix are related to the elements of the $M$ matrix as follows:

$$M'_{i,1} = M_{i,1} - M_{i,10}, \tag{39}$$

$$M'_{i,2} = M_{i,2} - M_{i,11}, \tag{40}$$

$$M'_{i,3} = M_{i,3} - M_{i,12}, \tag{41}$$

with $i$ ranging from 1 to 9. The rest of the elements in the $M'$ matrix are the same as those with the same indices in the $M$ matrix. The elements of the $B'$ vector are found, using Eq. (7), to be:

$$B' = -[M_{1,10}, M_{2,10}, M_{3,10}, M_{4,10}, M_{5,10}, M_{6,10}, M_{7,10}, M_{8,10}, M_{9,10}]^T. \tag{42}$$

From Eq. (12), it follows that:

$$A' = (M')^{-1}B'. \tag{43}$$

The $A$ vector can be then calculated from the $A'$ vector using Eq. (7). Specifically, the $A$ vector can be written as:

$$A = [A'_1, A'_2, A'_3, A'_4, A'_5, A'_6, A'_7, A'_8, A'_9, (1-A'_1), (-A'_2), (-A'_3)]^T. \tag{44}$$

The extension of this approach to arbitrary number of energy levels and keeping arbitrary order of time oscillating terms are described in **Appendix A**.

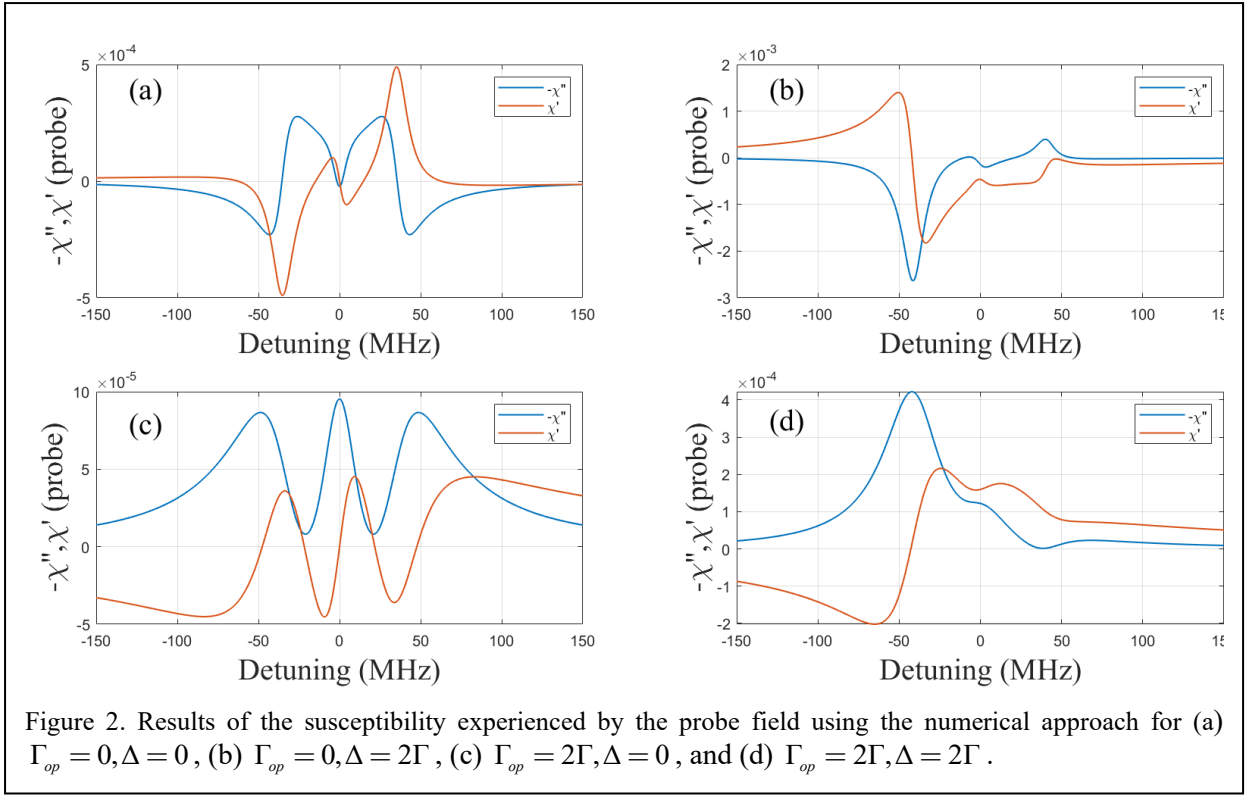

Figure 2. Results of the susceptibility experienced by the probe field using the numerical approach for (a) $\Gamma_{op}=0, \Delta=0$, (b) $\Gamma_{op}=0, \Delta=2\Gamma$, (c) $\Gamma_{op}=2\Gamma, \Delta=0$, and (d) $\Gamma_{op}=2\Gamma, \Delta=2\Gamma$.

This model can, of course, be used to derive the expectation value of any quantity of interest for the two level system. Specifically, we consider the susceptibility experienced by the probe field, which can be expressed as:

$$\chi = \frac{\hbar c_0 n_0}{I_{sat}\Omega_s}\left(\frac{\Gamma}{2}\right)^2 \tilde{\rho}_{21}^{-1}, \tag{45}$$

where $c_0$ is the speed of light in vacuum, $n_0$ is the number density of the atom, and $I_{sat}$ is the saturation intensity. To check the accuracy of this automated technique, it is instructive to compare the results produced by this model with results produced using explicit analysis of a two-level system under pump-probe excitations. The simplest case would be those studied in Refs. [2,3], in which there is no optical pumping from the ground state to the excited state. Since our model is more general, taking into account the possible presence of optical pumping from the ground to the excited state, we have chosen instead to consider the case studied earlier by us in Ref. 8, which considers the presence of such optical pumping. Figure 2 shows the real and imaginary parts of the susceptibility, for different combinations of pump detunings and optical pumping rates, corresponding to those shown in Figure 4 of Ref. [8]. The parameters used in these plots are as follows: $\Gamma = 2\pi \times 10^7 s^{-1}$, $\Omega_p = 2\pi \times 36 \times 10^6 s^{-1}$, $\Omega_s = 2\pi \times 6 \times 10^6 s^{-1}$, $I_{sat} = 120 W/m^2$, and $n_0 = 3 \times 10^{18} m^{-3}$. As can be seen, the ratios of the real and imaginary parts in each of these plots agree with the same found in Ref. [8]. The differences in the vertical scales are due to the use of different values of the dipole moments and number densities.

## 3. Symbolic approach

We have also developed another approach for solving the pump-probe model for an arbitrary number of levels, and to any order. This approach imitates the analytic solution in a code based on symbols and a set of MATLAB functions. We first present the basic concept for the case of a two-level system, keeping only the first order harmonics, and show how to generalize it to an arbitrary number of levels and orders later on in this section.

We start by defining symbols for all the density matrix elements, $\tilde{\rho}_{ij}^{0,\pm 1}$. We also define two additional symbols: $Y \equiv e^{i\delta t}$ and $Z \equiv e^{-i\delta t}$. Next, we evaluate the right-hand side of Eq. (6) by multiplying the matrices and adding the source term accordingly. We define the matrix $R$ as follows:

$$-\frac{i}{\hbar}\left(\tilde{\tilde{H}}\tilde{\rho} - \tilde{\rho}\tilde{\tilde{H}}^{\dagger}\right) + \tilde{\rho}_s \equiv R = \begin{bmatrix} R_{11} & R_{12} \\ R_{21} & R_{22} \end{bmatrix}. \tag{46}$$

For convenience, we vectorize the $R$ matrix in the form $V \equiv (R_{11}, R_{12}, R_{21}, R_{22})^{\mathrm{T}}$. The derivatives of the density matrix elements can be written as follows:

$$\dot{\tilde{\rho}}_{ij} = i\delta \tilde{\rho}_{ij}^{1} e^{i\delta t} - i\delta \tilde{\rho}_{ij}^{-1} e^{-i\delta t}. \tag{47}$$

We further define the following vectors:

$$\rho^P \equiv \left(\tilde{\rho}_{11}^1, \tilde{\rho}_{12}^1, \tilde{\rho}_{21}^1, \tilde{\rho}_{22}^1\right)^{\mathrm{T}}, \tag{48}$$

$$\rho^M \equiv \left(\tilde{\rho}_{11}^{-1}, \tilde{\rho}_{12}^{-1}, \tilde{\rho}_{21}^{-1}, \tilde{\rho}_{22}^{-1}\right)^{\mathrm{T}}. \tag{49}$$

We now define the following expressions:

$$Eq_n \equiv V_n - \left(i\delta \rho_n^P Y - i\delta \rho_n^M Z\right), \tag{50}$$

where $n = 1, 2, 3, 4$. Eq. (6) then corresponds simply to setting each of the expressions in Eq. (50) to zero. We note that each of these expressions would contain terms proportional to $YZ$, which is time independent, equaling unity. These would also contain terms proportional to $Y^2$ and $Z^2$, which represent higher order harmonics, and can be ignored. Next, for each of these expression, we generate three groups: one containing all terms that are time independent, one containing all terms proportional to $Y$, and one containing all terms proportional to $Z$. Each of these groups is then set to zero, which follows from the fact that we are considering the pseudo-steady state.

To implement the separation of these groups automatically, we make use of the symbolic MATLAB function $coeffs(Expression, [Symbol\,1, Symbol\,2, \cdots, Symbol\,n])$. Briefly, this function takes as inputs an expression and $n$ symbols, and returns two output vectors, which can be best illustrated using an example, as follows. Assume we have an expression defined as $eq = ax^2 + bxy + cx + dy^2 + ey + f$, where $x$ and $y$ are the symbols of interest. The command $[p,q] = coeffs(eq, [x, y])$ returns two vectors: the vector $p$ would be $p = [x^2, xy, x, y^2, y, 1]$ and the corresponding vector $g$ would be $q = [a, b, c, d, e, f]$.

When provided with $Eq_n$ as an input expression, along with the relevant variables, which are $Y$ and $Z$, this function returns the coefficients in $Eq_n$ for each variable, i.e., $Y$, $Z$, and any combinations thereof. For each expression, $eq$, the MATLAB function $[g,h] = coeffs(eq,[Y,Z])$ returns two vectors, $g$ and $h$. The vector $h$ contains terms $[Y^2, YZ, Y, Z^2, Z, 1]$, and the corresponding vector $g$ contains terms we denote as $[q,r,s,u,v,w]$. Here we ignore the coefficients for $Y^2$ and $Z^2$. Since $Y \cdot Z = 1$, the coefficients of $YZ$ and 1 are grouped in the same expression. The expression $eq$ can then be separated into three expressions: $eq_a = r+w$, $eq_b = s$, and $eq_3 = v$. This process leads to a set of 12 linear expressions, each of which is equated to zero. Together, these equations correspond to Eq. (8).

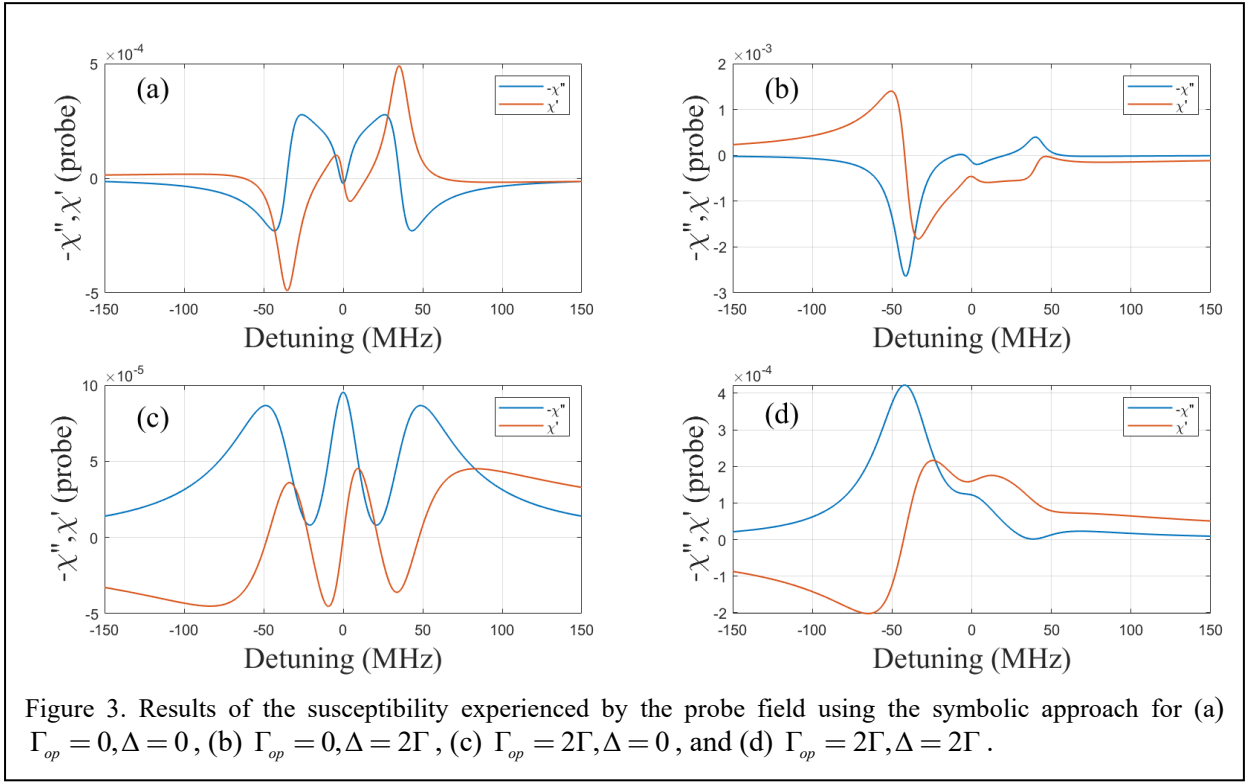

Figure 3. Results of the susceptibility experienced by the probe field using the symbolic approach for (a) $\Gamma_{op} = 0, \Delta = 0$, (b) $\Gamma_{op} = 0, \Delta = 2\Gamma$, (c) $\Gamma_{op} = 2\Gamma, \Delta = 0$, and (d) $\Gamma_{op} = 2\Gamma, \Delta = 2\Gamma$.

To convert the set of equations to a matrix notation, one can use another symbolic MATLAB function $[M,B] = equationsToMatrix(EquationVector, VariableVector)$. It takes as an input a set of $m$ equations organized as a column vector, and a set of $m$ variables organized as a column vector. The out M is an $m \times m$ matrix, and the output B is an $m \times 1$ column vector. In our case, the value of $m$ is 12, the EquationVector is $[eq_1, eq_2 \cdots, eq_{12}]^T$, and the VariableVector is $[\tilde{\rho}_{11}^0, \tilde{\rho}_{11}^1, \tilde{\rho}_{11}^{-1}, \tilde{\rho}_{12}^0, \cdots, \tilde{\rho}_{22}^{-1}]^T$, which is the same as the column vector $A$ shown earlier in Eq. (9), using the same rule for ordering. This function returns the coefficient matrix $M$ and the $B$ vector satisfying Eq. (8). Next, we apply the constraints for a closed system and repeat the steps described in Eq. (39) to Eq. (44) for the numerical approach to determine the pseudo-steady state solution for the density matrix equations of motion.

We next discuss how to generalize the symbolic approach. Consider first the process of generalization to an arbitrary number of energy levels, while keeping only the first order terms. This simply requires changing the length of the vectors defined above based on the number of levels, with the same rule for ordering. To be specific, the length of vectors $V$, $\rho^P$, and $\rho^M$ for $N$ energy levels is $N^2$ for keeping only the first order harmonics. The procedure for separating the linear equations and solving the density matrix remains the same. Finally, we have to apply the constraints for a closed system to determine the pseudo-steady state solution for the density matrix equations of motion. Just as in the case of the two-level system, the steps needed for this are identical to those used for the numerical approach for $N$ energy levels, which are described in Eqs. (64) to (72) in Appendix A.

Expanding the derivation to the $K$-th order requires taking the coefficients up to $Y^K$ and $Z^K$ and the relevant combinations of parameters. It needs to be noted that the Hamiltonian only has the first order harmonics. As a result, the pseudo-commutator will only have the factors in the form of $Y^m$, $Z^m$, $Y^m Z$, and $YZ^m$ with $1 < m < K$. It can be seen that $Y^m Z$ need to be grouped with $Y^{m-1}$ and $YZ^m$ need to be grouped with $Z^{m-1}$, which leads to the fact that the coefficients of these terms should be added together when determining the corresponding equations. Moreover, the length of the $A$ vector would be $(2K+1)N^2$.

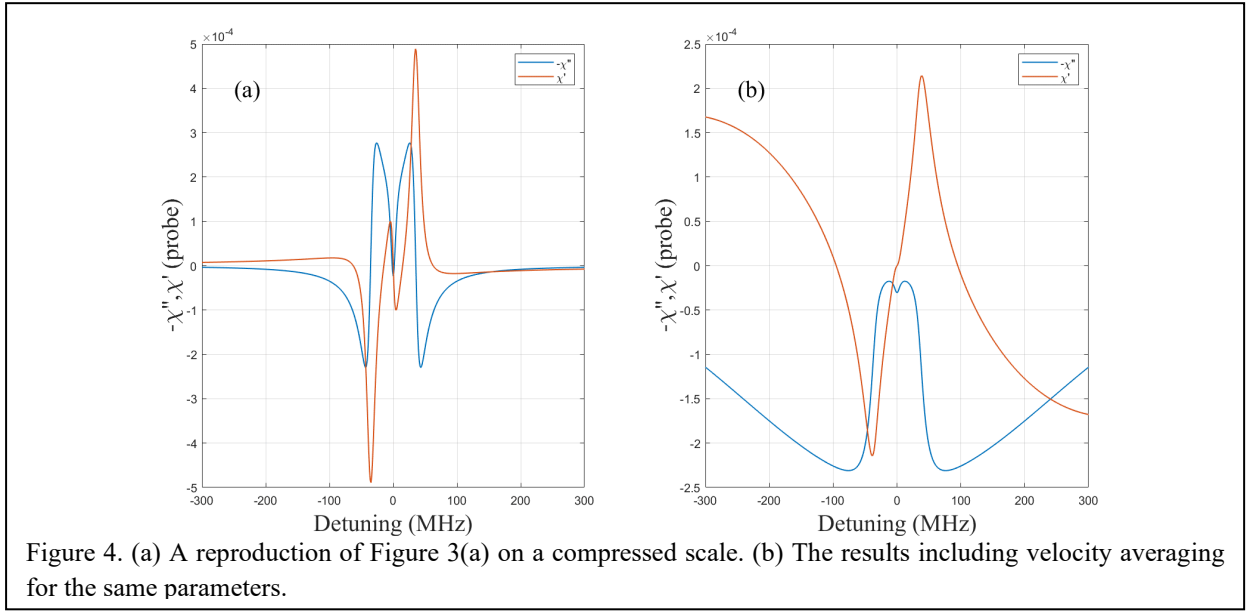

Figure 4. (a) A reproduction of Figure 3(a) on a compressed scale. (b) The results including velocity averaging for the same parameters.

Figure 3 shows the real and imaginary parts of the probe susceptibility, as functions of its detuning. As can be seen, these agree exactly with the results shown in Figure 2, produced using the numerical approach. All the simulation results shown in the rest of this paper have been carried out with both approaches, yielding the same results.

In principle, one could use either the numerical approach or the symbolic approach. However, we have found that the amount of time needed to carry out the computation is significantly larger for the symbolic approach, even for a two-level system up to the first order, and the difference in the computation speed grows with increasing number of energy levels and orders. As such, in practice, one should use the numerical approach. On the other hand, it should be noted that while generalizing to arbitrary numbers of levels and arbitrary orders requires complex steps requiring close attention for the numerical case, it is much simpler for the symbolic case. As such, when creating new codes under such scenarios, one should use both approaches first to make sure they produce identical results, and then use the numerical codes for investigating the behavior of the system as a function of various parameters, as well as for velocity averaging.

We have studied the susceptibility dependence on the probe detuning, for a Doppler width of 564 MHz (FWHM), as illustrated in Figure 4. Figure 4(a) is a reproduction of Figure 3(a) with a different horizontal scale, and Figure 4(b) shows the results when including velocity averaging for the same set of parameters.

In the simulation results shown so far, we only considered the case of a vanishingly weak probe. However, there can be many situations in which this limit does not hold. For such cases, one must keep adding effects of higher orders, until the addition of one more order produces changes in the result deemed negligible for the application at hand.

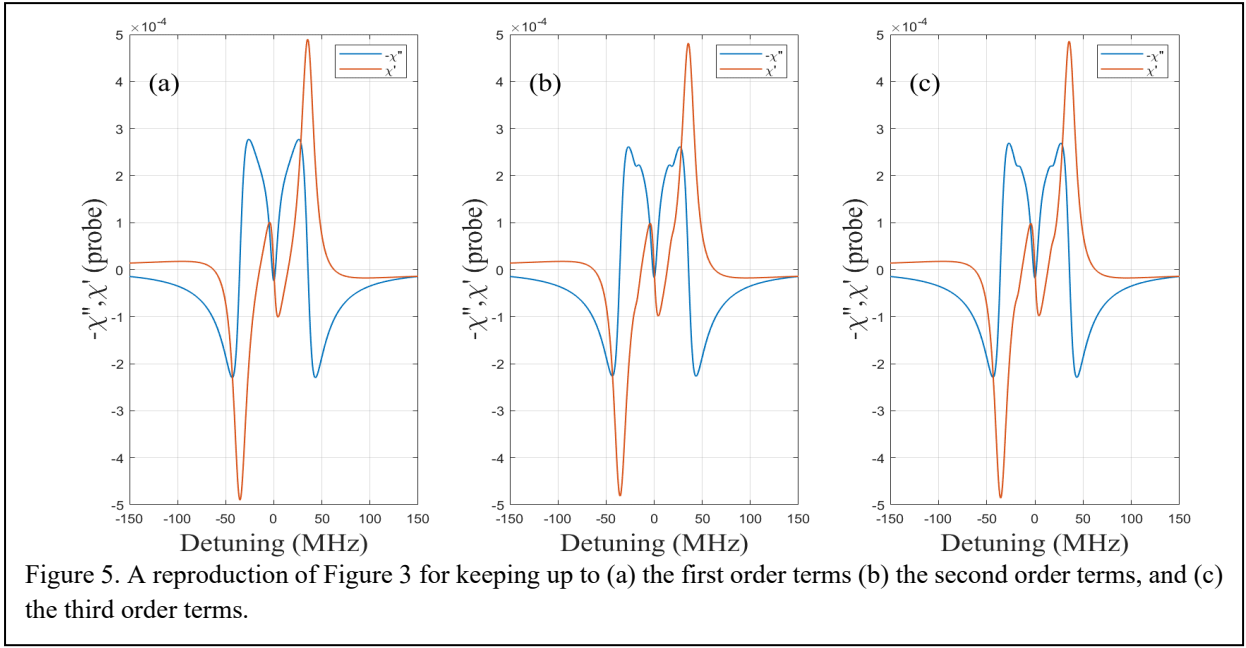

Figure 5. A reproduction of Figure 3 for keeping up to (a) the first order terms (b) the second order terms, and (c) the third order terms.

Figure 5 shows the simulation results, for the parameters used in Figure 3(a), when the maximum order used is one (a), 2(b) and 3(c). As can be seen, minor deviations exist between the cases, especially around the detuning of $\pm 18$ MHz.

## 4. Three-level Lambda system

We have also studied the case of a 3-level Lambda system where two different frequencies are applied along one of the two legs, as shown in Figure 6. This case represents an idealized version of the scheme for realizing a superluminal laser using a single pump [13]. Here, the pump couples two transitions: the transition $|2\rangle \leftrightarrow |3\rangle$ with detuning $\delta_{p23}$ and the transition $|1\rangle \leftrightarrow |3\rangle$ with detuning $\delta_{p13}$. The two detunings of the Raman pump are related as $\delta_{p13} \equiv \delta_{p23} - \Delta$, where $\Delta$ is the hyperfine ground states separation. For simplicity of notation, we defined $\delta_p \equiv \delta_{p23}$. A probe beam is applied along the $|2\rangle \leftrightarrow |3\rangle$ transition with detuning $\delta_s$. The Hamiltonian representing the system under the RWA and the RWT is presented as:

$$\tilde{\tilde{H}} \equiv \frac{\hbar}{2} \begin{bmatrix} -i\Gamma_g - 2\Delta & 0 & \Omega_p \\ 0 & -i\Gamma_g & \Omega_p + \Omega_s e^{i\delta t} \\ \Omega_p & \Omega_p + \Omega_s e^{-i\delta t} & -i\Gamma - 2\delta_p \end{bmatrix}, \quad (51)$$

where $\Gamma_g$ is the collisional decay rate and $\delta \equiv \delta_s - \delta_p$. The source term can be expressed as:

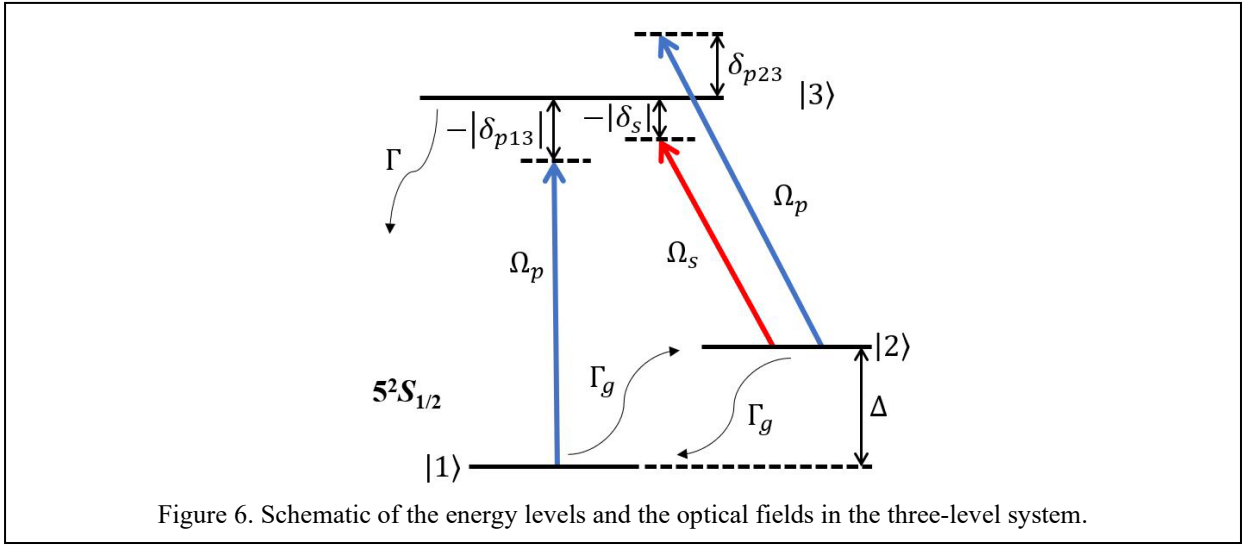

Figure 6. Schematic of the energy levels and the optical fields in the three-level system.

$$\tilde{\rho}_{source} \equiv \begin{bmatrix} \Gamma_g \tilde{\rho}_{22} + \Gamma \tilde{\rho}_{33}/2 & 0 & 0 \\ 0 & \Gamma_g \tilde{\rho}_{11} + \Gamma \tilde{\rho}_{33}/2 & 0 \\ 0 & 0 & 0 \end{bmatrix}. \qquad (52)$$

The $A$ vector can be written in the same form as Eq. (9):

$$A \equiv \left[ \tilde{\rho}_{11}^0, \tilde{\rho}_{11}^1, \tilde{\rho}_{11}^{-1}, \tilde{\rho}_{12}^0, \tilde{\rho}_{12}^1, \tilde{\rho}_{12}^{-1}, \tilde{\rho}_{13}^0, ..., \tilde{\rho}_{33}^0, \tilde{\rho}_{33}^1, \tilde{\rho}_{33}^{-1} \right]^T. \qquad (53)$$

Applying the algorithms described above to that system yields the results shown in Figure 7. We observe the Autler Towns splitting [34] and the corresponding negative dispersion slope between the peaks. The results show the population inversion between level 1 and level 2, as expected.

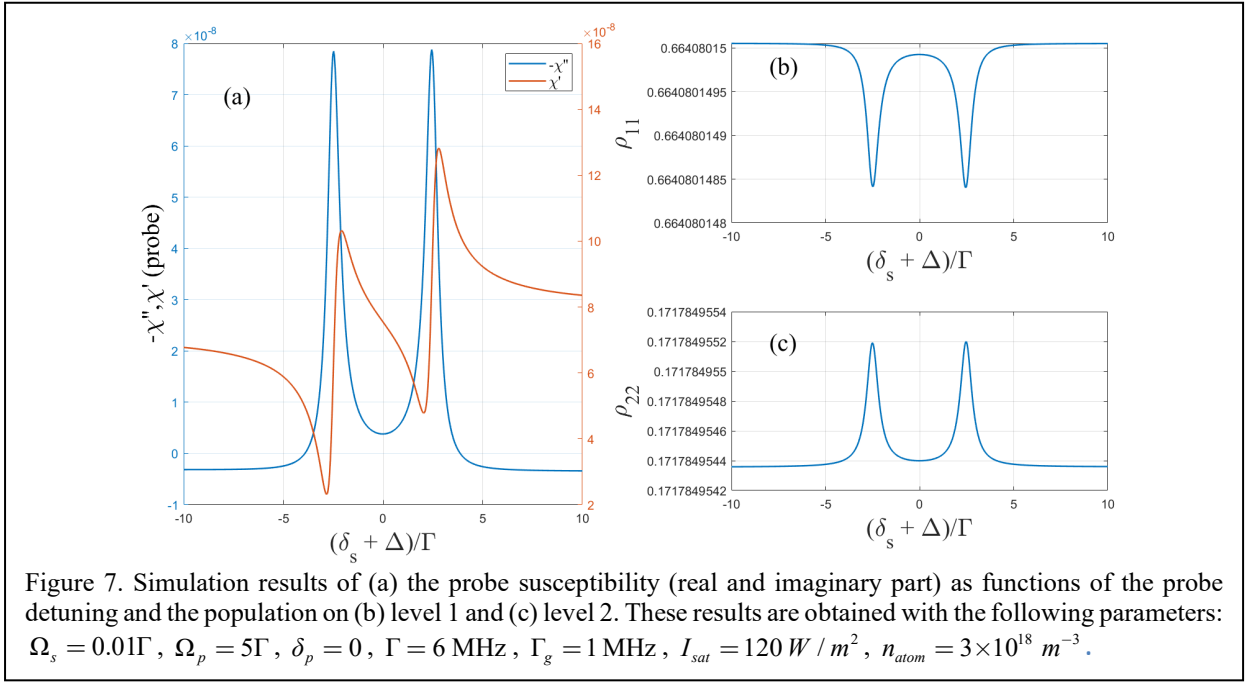

Figure 7. Simulation results of (a) the probe susceptibility (real and imaginary part) as functions of the probe detuning and the population on (b) level 1 and (c) level 2. These results are obtained with the following parameters: $\Omega_s = 0.01\Gamma$, $\Omega_p = 5\Gamma$, $\delta_p = 0$, $\Gamma = 6\,\text{MHz}$, $\Gamma_g = 1\,\text{MHz}$, $I_{sat} = 120\,W/m^2$, $n_{atom} = 3 \times 10^{18}\,m^{-3}$.

In the preceding discussion, we used two and three-level systems to illustrate the application of these algorithms. However, as noted earlier, these algorithms work for an arbitrary number of levels. We next show another example based on a four-level model.

## 5. Four-level system as a model for the single-pumped superluminal laser

The four-level model described in Figure 8 is similar to the three-level model, but with an additional excited state denoted as level $|4\rangle$. The two excited states, separated by $\Delta_{23}$, represent the hyperfine structure of the $5P_{1/2}$ manifold. Here, both the pump (blue arrow) and the probe (red arrow) couple level $|3\rangle$ and level $|4\rangle$ to level $|2\rangle$. In principle, both the pump and the probe also couple level $|3\rangle$ and level $|4\rangle$ to level $|1\rangle$. However, since the splitting between level $|1\rangle$ and $|2\rangle$ is very large (~3 GHz) compared to the Doppler broadening (~600 MHz) and the probe Rabi frequency even when the system is lasing at the probe frequency, it is reasonable to assume that the coupling of levels $|3\rangle$ and level $|4\rangle$ to level $|1\rangle$ due to the probe is negligible. As such, this configuration represents the system used for the single-pumped superluminal laser scheme described in Ref. [13]. In what follows, we apply the pump-probe algorithm to this system to generate the gain spectrum for the probe, under idealized conditions. We will then discuss how to augment this model to take into account all the Zeeman sublevels in order to yield more accurate results.

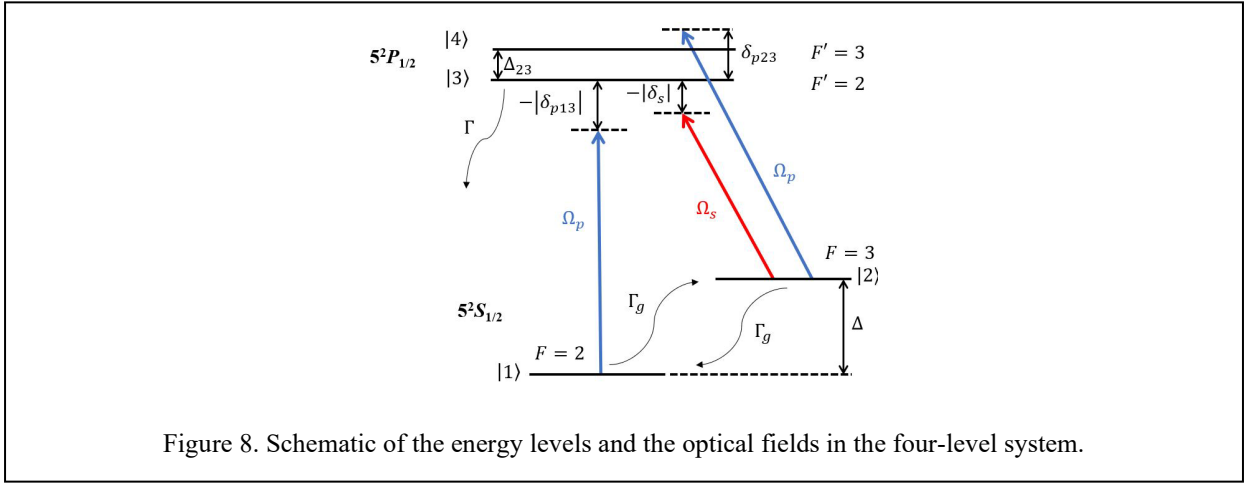

Figure 8. Schematic of the energy levels and the optical fields in the four-level system.

To start with, we assume that the pump Rabi frequencies for the $|1\rangle \leftrightarrow |3\rangle$, $|1\rangle \leftrightarrow |4\rangle$, $|2\rangle \leftrightarrow |3\rangle$ and $|2\rangle \leftrightarrow |4\rangle$ transitions are identical, and given by $\Omega_P$. Similarly, we assume that the probe Rabi frequencies for the $|2\rangle \leftrightarrow |3\rangle$ and $|2\rangle \leftrightarrow |4\rangle$ transitions are identical, and given by $\Omega_S$. All the detunings are defined in the same way as for the three level system described above. The four-level Hamiltonian under the RWA and the RWT can then be written as:

$$\tilde{\tilde{H}} \equiv \frac{\hbar}{2}\begin{bmatrix} -i\Gamma_g - 2\Delta & 0 & \Omega_p & \Omega_p \\ 0 & -i\Gamma_g & \Omega_p + \Omega_s e^{i\delta t} & \Omega_p + \Omega_s e^{i\delta t} \\ \Omega_p & \Omega_p + \Omega_s e^{-i\delta t} & -i\Gamma - 2\delta_p & 0 \\ \Omega_p & \Omega_p + \Omega_s e^{-i\delta t} & 0 & -i\Gamma - 2\delta_p + 2\Delta_{23} \end{bmatrix}. \quad (54)$$

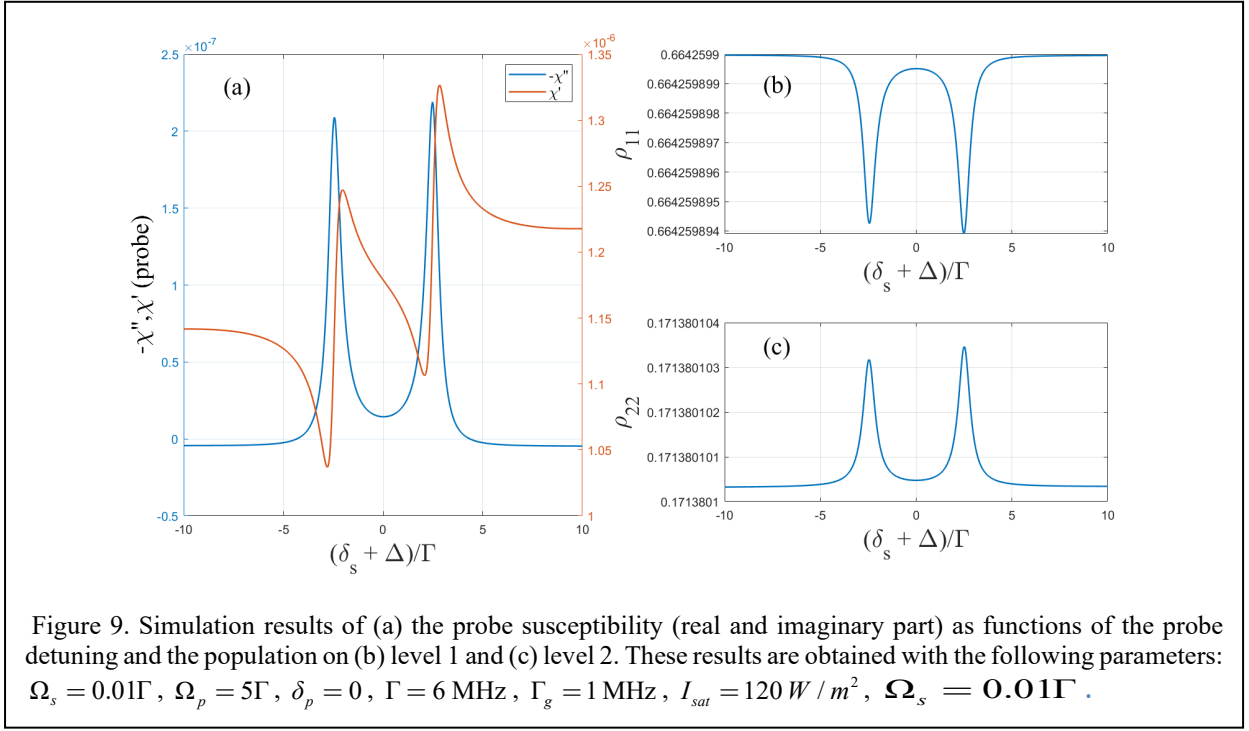

Figure 9. Simulation results of (a) the probe susceptibility (real and imaginary part) as functions of the probe detuning and the population on (b) level 1 and (c) level 2. These results are obtained with the following parameters: $\Omega_s = 0.01\Gamma$, $\Omega_p = 5\Gamma$, $\delta_p = 0$, $\Gamma = 6$ MHz, $\Gamma_g = 1$ MHz, $I_{sat} = 120\ W/m^2$, $\Omega_s = 0.01\Gamma$.

The source term can be expressed as:

$$\tilde{\rho}_{source} \equiv \begin{bmatrix} \Gamma_g \tilde{\rho}_{22} + \Gamma\tilde{\rho}_{33}/2 + \Gamma\tilde{\rho}_{44}/2 & 0 & 0 & 0 \\ 0 & \Gamma_g \tilde{\rho}_{11} + \Gamma\tilde{\rho}_{33}/2 + \Gamma\tilde{\rho}_{44}/2 & 0 & 0 \\ 0 & 0 & 0 & 0 \\ 0 & 0 & 0 & 0 \end{bmatrix}. \quad (55)$$

Figure 9 shows the simulation results under the same parameters as those used for the three-level case. Here, we observe asymmetry between the peaks in the susceptibility and the populations due to the coupling to an additional excited state. To illustrate the symmetric splitting, which is the case of interest for the single-pumped superluminal laser [13], one would have to modify the pump detuning accordingly.

While this approach is expected to be more accurate than the model employed in Ref. [13], the idealized results shown in Figure 9 are only illustrative, and not expected to correspond to experimentally observed results. This is because a comprehensive model has to take into account the Zeeman sublevels within the hyperfine states, which adds up to a 24-level system. Determining experimentally verifiable gain profile and sensitivity of the resulting superluminal laser would require accounting for all these Zeeman sublevels, keeping track of many harmonic terms since the field inside the laser cannot be treated as weak, velocity averaging, and iterative solution of the laser equations in a self-consistent manner. Such an investigation, which is extremely time consuming, even with a supercomputer, is currently in progress, and the results would be reported in the near future.

## 6. 16-level system for the self-pumped Raman gain

Finally, to demonstrate that these algorithms are capable of calculating a complex system without applying the approximations that circumvent the pump-probe issue, we calculate the self-pumped Raman gain produced in the D1 line in $^{87}$Rb. Here we consider all the Zeeman sublevels in the relevant energy levels for this system, in total 16 levels, as shown in Figure 10(a). The hyperfine splitting between the states $5S_{1/2}$, F=1 and $5S_{1/2}$, F=2 is ~6.835 GHz. Each Zeeman sublevel in the $5P_{1/2}$ manifold decays to the $5S_{1/2}$ manifold, at a rate denoted as $\Gamma$. The branching ratio for the decay to each Zeeman sublevel in the $5P_{1/2}$ manifold is different and determined by the dipole matrix elements. Between each pair of the Zeeman sublevels, when one is from F=1 and the other is from F=2, in the $5S_{1/2}$ manifold, there is a collisional decay rate of 1 MHz.

The pump field and the probe field are cross-linearly polarized, which determines the signs of the Rabi frequencies of the two fields and the expression for calculating the susceptibility. The pump field couples all the allowed Zeeman transitions in the D1 manifold, specifically the $5S_{1/2}$, F=1 to $5P_{1/2}$, F=1 and $5P_{1/2}$, F=2 transitions, and the $5S_{1/2}$, F=2 to $5P_{1/2}$, F=1 and $5P_{1/2}$, F=2 transitions . The probe field couples the $5P_{1/2}$, F=1 and $5P_{1/2}$, F=2 transitions. It needs to be noted that only the $\sigma^+$ fields are shown in the diagram for clarity. In the complete system both $\sigma^+$ fields and $\sigma^-$ fields are considered. The pump field is tuned near the resonance of the $5S_{1/2}$, F=2 to $5P_{1/2}$ transitions. As such, on the $5S_{1/2}$, F=1 to $5P_{1/2}$ transitions, the pump field is detuned below resonance by ~6.835 GHz. The Rabi frequency of each Zeeman transition is related to those of the other Zeeman transitions by the corresponding matrix element of the dipole moment [35]. The Doppler broadening is taken into account by averaging the density matrix over several different velocity groups with the thermal distribution at 100 °C.

The gain and dispersion spectra experienced by the probe field, which are determined by adding contributions from all the probe Zeeman transitions, are shown in Figure 10(b). The center of the x-axis corresponds to the two-photon detuning being zero. The result is generated with the pump field detuned by $30\Gamma$ with respect to the $5S_{1/2}$, F=2 to $5P_{1/2}$, F=2 transition. The Rabi frequencies of the pump field are evaluated by multiplying the dipole matrix elements by $10\Gamma$. As can be seen, the peak gain experienced by the probe field is above unity, which can be used for producing self-pumped Raman lasing.

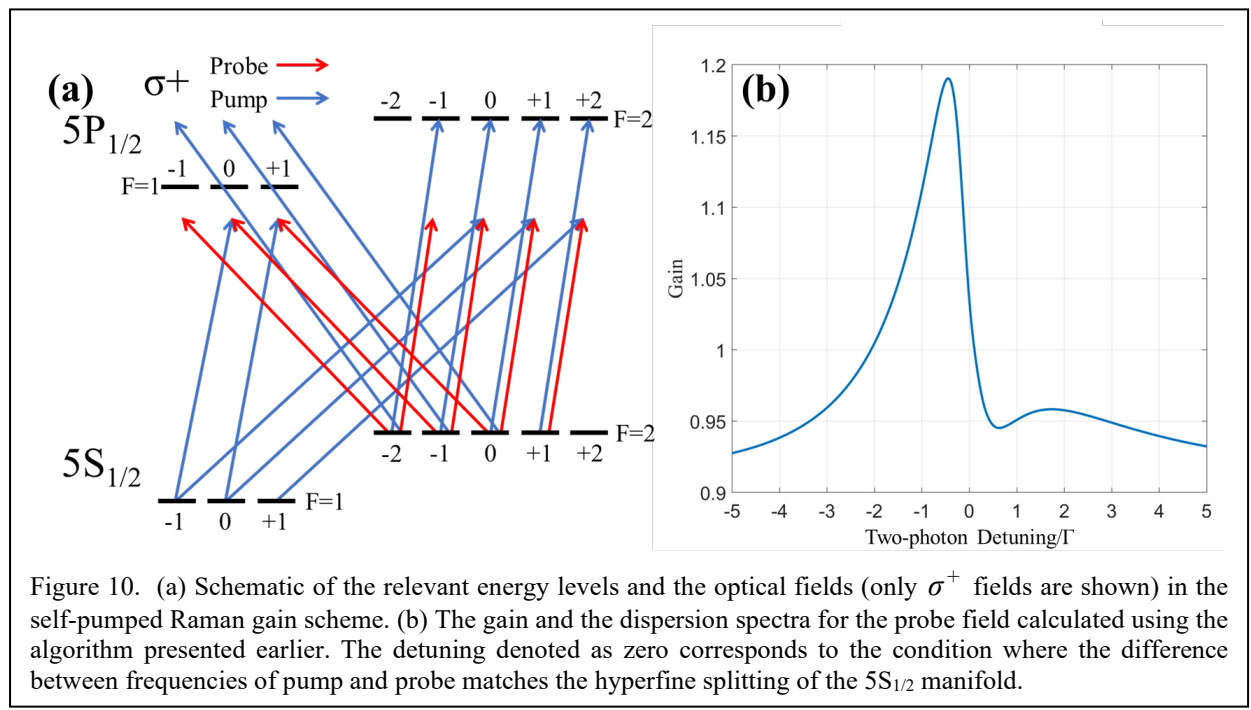

Figure 10. (a) Schematic of the relevant energy levels and the optical fields (only $\sigma^+$ fields are shown) in the self-pumped Raman gain scheme. (b) The gain and the dispersion spectra for the probe field calculated using the algorithm presented earlier. The detuning denoted as zero corresponds to the condition where the difference between frequencies of pump and probe matches the hyperfine splitting of the $5S_{1/2}$ manifold.

In **Appendix B**, we show the MATLAB code for implementing the numerical algorithm, and in **Appendix C**, we show the MATLAB code for implementing the symbolic algorithm.

## 7. Discussions and Conclusions

In this paper, we describe a generalized algorithm for evaluating the steady-state solution of the density matrix equation of motion, for situations where two fields oscillating at different frequencies couple the same set of atomic transitions involving an arbitrary number of energy levels, to an arbitrary order of the harmonics of the frequency difference between the pump and the probe. We developed a numerical approach as well as a symbolic one for implementing this algorithm. We have verified that both approaches yield the same results for all cases studied. However, we have found that the numerical approach is significantly faster. On the other hand, generalizing to arbitrary numbers of levels and arbitrary orders requires complex steps for the numerical case, but is much simpler for the symbolic case. As such, when creating new codes under such scenarios, one should use both approaches first to make sure they produce identical results, and then use the numerical codes for investigating the behavior of the system as a function of various parameters. The validity of the codes has been established by comparing them with the analytical solution of a two-level system to first order. We have also produced results up to the third order harmonics for a two-level system, and to first order for three- and four-level systems,. In addition, we have used this model to accurately determine the gain profile for a self-pumped Raman laser in the D1 manifold of $^{87}$Rb atoms, taking into account all 16 Zeeman sublevels. By eliminating the need for making the approximation that each field couples only one transition, this algorithm can yield accurate results of numerical calculations for many practical systems of interest.

**Funding.** This work has been supported by AFOSR (FA9550-18-01-0401, FA9550-21-C-0003 and FA9550-23-1-0617), NASA (80NSSC22CA052), Defense Security Cooperation Agency (PO4441028735), Israeli MOD (4441185451), and Israeli Innovation Authority (4692/1).
**Disclosure.** The authors declare no conflicts of interests.
**Data availability.** Data underlying the results presented in this paper are not publicly available at this time but may be obtained from the authors upon reasonable request.

## Appendix A: Extension of the numerical approach to an arbitrary number of energy levels and keeping arbitrary orders of time oscillating terms

In this appendix, we describe the generalization of the numerical approach for a system with an arbitrary number of energy levels, while keeping arbitrary orders of time oscillating terms. For the sake of clarity, we follow a two-step process in this description. In Section A.1, we describe the process for applying the numerical approach to an arbitrary number of energy levels, while keeping only the first order terms. In Section A.2, we present the process for keeping arbitrary orders of time oscillating terms, for an arbitrary number of energy levels.

### A.1. Extending the numerical approach to an arbitrary number of energy levels

Here, we describe the process for generalizing the numerical algorithm for an arbitrary number of energy levels. In order to keep the description simple, we only keep up to the first order terms, i.e. the terms that are multiplied by $e^{\pm i\delta t}$. The more general case which involves an arbitrary number of energy levels while keeping terms to arbitrary orders is presented in the next section.

We first write the $A$ vector in the same form as that in Eq. (9):

$$A \equiv \left[\tilde{\rho}_{11}^0, \tilde{\rho}_{11}^1, \tilde{\rho}_{11}^{-1}, \tilde{\rho}_{12}^0, \tilde{\rho}_{12}^1, \tilde{\rho}_{12}^{-1}, \tilde{\rho}_{13}^0, ..., \tilde{\rho}_{NN}^0, \tilde{\rho}_{NN}^1, \tilde{\rho}_{NN}^{-1}\right]^T. \tag{56}$$

For a system with $N$ energy levels, $A$ is a $3N^2 \times 1$ column vector. The ordering for the linear equations would follow the same rule as that for the elements in the $A$ vector.

The process for evaluating the $M$ matrix, with a dimension of $3N^2 \times 3N^2$, remains the same as that for the two-level system. Specifically, we start with regrouping the linear equations. The steps described from Eq. (17) to Eq. (27) are valid for a system with N energy levels except that each of the matrices ($\tilde{H}$, $\tilde{\rho}$, $U$, and $G$) has a dimension of $N \times N$. As a result, the total number of the linear equations is $3N^2$. Each of $G^0$, $G^1$, and $G^{-1}$ contains $N^2$ equations that are time independent. We then set one of the elements in the $A$ vector to unity and the other elements to zeroes to find the corresponding coefficient in the $M$ matrix. The pseudo-commutator is evaluated, along with proper addition of source terms. When setting $v$-th element in the $A$ vector to be unity and others to null values, we first need to figure out the row and the column number of this element in the density matrix. Here we define that the only non-zero element in the $A$ vector is located at the $i$-th row and $j$-th column in the density matrix. We then have the relations: $i = \lceil v/(3N) \rceil$ and $j = \lceil [v - 3N(i-1)]/3 \rceil$ ($\lceil \ \rceil$: round toward positive infinity). Here, we define an $N \times N$ matrix, $\Lambda$, where all elements have null values except for the element at the $i$-th row and $j$-th column, which has a value of unity:

$$\Lambda_{pq} = \begin{cases} 1, & p = i \text{ and } q = j \\ 0, & p \neq i \text{ or } q \neq j \end{cases}. \tag{57}$$

The three unique matrices for the pseudo-commutator are:

$$Q^0 \equiv H^0 \Lambda - \Lambda \left(H^0\right)^\dagger; \tag{58}$$

$$Q^1 \equiv H^1 \Lambda - \Lambda H^1; \tag{59}$$

$$Q^{-1} \equiv H^{-1} \Lambda - \Lambda H^{-1}. \tag{60}$$

The linear equation used for determining the $u$-th row and the $v$-th column element in the $M$ matrix is:

$$E_{u(LHS)} = G_{xy}^z = -\frac{i}{\hbar}U_{xy}^z + \tilde{\rho}_{s(xy)}^z. \tag{61}$$

where the subscripts $x$ and $y$ indicate the row and the column of the matrices, respectively, which can be determined using $x = \lceil u/(3N) \rceil$ and $y = \lceil [u-3N(i-1)]/3 \rceil$; and $z$ is the superscript for the $u$-th linear equation which can be determined using the following rule:

$$z = \begin{cases} 0, & \text{if } u - 3N(x-1) - 3(y-1) = 0 \\ 1, & \text{if } u - 3N(x-1) - 3(y-1) = 1 \\ -1, & \text{if } u - 3N(x-1) - 3(y-1) = 2 \end{cases}. \tag{62}$$

The term $U_{xy}^z$ can be evaluated using Eq. (21) to Eq. (23) and Eq. (58) to Eq. (60). Then we can find the value of $M_{uv}$ as:

$$M_{uv} = E_{u(LHS)}\big|_{\Lambda_{ij}=1}. \tag{63}$$

All elements of the $M$ matrix can be determined by repeating this process.

Once the $M$ matrix is evaluated, we apply the closed system constraints to remove the redundant equations. Slightly different from the two-level system, the closed system constraints for an N-level system become:

$$\sum_{i=1}^{N} \tilde{\rho}_{ii}^0 = 1, \quad \sum_{i=1}^{N} \tilde{\rho}_{ii}^1 = 0, \quad \sum_{i=1}^{N} \tilde{\rho}_{ii}^{-1} = 0. \tag{64}$$

As a result, Eq. (39) to Eq. (41) need to be modified. For an arbitrary number of energy levels, $N$, the columns in the $M'$ matrix that correspond to $\tilde{\rho}_{ii}^0$, $\tilde{\rho}_{ii}^1$, and $\tilde{\rho}_{ii}^{-1}$ with $i = 1, 2, ..., (N-1)$, needs to be modified due to the closed system constraints while the rest of the columns are unchanged. Specifically, several columns have to be modified according to the following equations:

$$M'_{j,[3(i-1)(N+1)+1]} = M_{j,[3(i-1)(N+1)+1]} - M_{j,3N^2-2}. \tag{65}$$

$$M'_{j,[3(i-1)(N+1)+2]} = M_{j,[3(i-1)(N+1)+2]} - M_{j,3N^2-1}. \tag{66}$$

$$M'_{j,[3(i-1)(N+1)+3]} = M_{j,[3(i-1)(N+1)+3]} - M_{j,3N^2}. \tag{67}$$

for $j = 1, 2, ..., 3N^2 - 3$ and $i = 1, 2, ..., N-1$.

The $B'$ vector can be written as:

$$B' \equiv -\left[M_{1,3N^2-2}, M_{2,3N^2-2}, ..., M_{3N^2-3,3N^2-2}\right]^T. \tag{68}$$

The evaluation of the $A'$ vector is done in the same way as in the case of the two-level system via Eq. (43). However, to determine the $A$ vector, the process is slightly different. To be more specific, the elements $\tilde{\rho}_{ii}^0$, $\tilde{\rho}_{ii}^1$, and $\tilde{\rho}_{ii}^{-1}$ with $i = 1, 2, ..., (N-1)$ need to be used to evaluate the last three elements in the $A$ vector:

$$A_{3N^2-2,1} = 1 - \sum_{i=1}^{N-1} \tilde{\rho}_{ii}^0, \quad A_{3N^2-1,1} = -\sum_{i=1}^{N-1} \tilde{\rho}_{ii}^1, \quad A_{3N^2,1} = -\sum_{i=1}^{N-1} \tilde{\rho}_{ii}^{-1}. \tag{69}$$

Equivalently, these terms can be expressed in terms of $A'$ vector elements as follows:

$$A_{3N^2-2,1} = 1 - \sum_{i=1}^{N-1} A'_{3(i-1)(N+1)+1,1}, \tag{70}$$

$$A_{3N^2-1,1} = -\sum_{i=1}^{N-1} A'_{3(i-1)(N+1)+2,1}, \tag{71}$$

$$A_{3N^2,1} = -\sum_{i=1}^{N-1} A'_{3(i-1)(N+1)+3,1}. \tag{72}$$

**A.2. Extend the numerical approach for keeping arbitrary orders of time oscillating terms**

As can be seen in Eqs. (17) to (23), when keeping terms up to the first order, the pseudo-commutator yields the products of terms with different superscripts. To eliminate the time dependent factors, it is necessary to group the terms that are multiplied by $e^{\pm i\delta t}$. In this case where we ignore the terms that are multiplied by $e^{\pm i2\delta t}$, it is straightforward to derive the expression for $U^0$, $U^1$, and $U^{-1}$. However, this process becomes somewhat more involved when keeping higher order terms, as illustrated below.

To generalize the numerical approach for keeping up to *K*-th order terms, we start with the definition of the density matrix and the source matrix:

$$\tilde{\rho} = \tilde{\rho}^0 + \tilde{\rho}^1 e^{i\delta t} + \tilde{\rho}^{-1} e^{-i\delta t} + \tilde{\rho}^2 e^{i2\delta t} + \tilde{\rho}^{-2} e^{-i2\delta t} + ... + \tilde{\rho}^K e^{iK\delta t} + \tilde{\rho}^{-K} e^{-iK\delta t}, \tag{73}$$

$$\tilde{\rho}_s = \tilde{\rho}_s^0 + \tilde{\rho}_s^1 e^{i\delta t} + \tilde{\rho}_s^{-1} e^{-i\delta t} + \tilde{\rho}_s^2 e^{i2\delta t} + \tilde{\rho}_s^{-2} e^{-i2\delta t} + ... + \tilde{\rho}_s^K e^{iK\delta t} + \tilde{\rho}_s^{-K} e^{-iK\delta t}. \tag{74}$$

The Liouville equation in pseudo steady state can be expressed as:

$$\frac{\partial \tilde{\rho}}{\partial t} = 0 + i\delta\tilde{\rho}^1 e^{i\delta t} - i\delta\tilde{\rho}^{-1} e^{-i\delta t} + i2\delta\tilde{\rho}^2 e^{i2\delta t} - i2\delta\tilde{\rho}^{-1} e^{-i2\delta t} + ... + iK\delta\tilde{\rho}^K e^{iK\delta t} - iK\delta\tilde{\rho}^{-K} e^{-iK\delta t}$$

$$= -\frac{i}{\hbar}\left[\tilde{\tilde{H}}\tilde{\rho} - \tilde{\rho}\tilde{\tilde{H}}^\dagger\right] + \tilde{\rho}_s. \tag{75}$$

It needs to be noted that the Hamiltonian remains the same as in Eqs. (13) to (16), generalized for an arbitrary number of energy levels. The pseudo-commutator can be written as:

$$\tilde{\tilde{H}}\tilde{\rho} - \tilde{\rho}\tilde{\tilde{H}}^\dagger = \left(H^0 + H^1 e^{i\delta t} + H^{-1} e^{-i\delta t}\right)\left(\tilde{\rho}^0 + \tilde{\rho}^1 e^{i\delta t} + \tilde{\rho}^{-1} e^{-i\delta t} + ... + \tilde{\rho}^K e^{iK\delta t} + \tilde{\rho}^{-K} e^{-iK\delta t}\right)$$
$$-\left(\tilde{\rho}^0 + \tilde{\rho}^1 e^{i\delta t} + \tilde{\rho}^{-1} e^{-i\delta t} + ... + \tilde{\rho}^K e^{iK\delta t} + \tilde{\rho}^{-K} e^{-iK\delta t}\right)\left(H^0 + H^1 e^{i\delta t} + H^{-1} e^{-i\delta t}\right)^\dagger. \tag{76}$$

We expand and rearrange the right hand side of Eq. (76) and define the following quantities with the approximation that the terms varying faster than $e^{\pm iK\delta t}$ (such as $e^{\pm i(K+1)\delta t}$) are dropped:

$$\tilde{\tilde{H}}\tilde{\rho} - \tilde{\rho}\tilde{\tilde{H}}^\dagger \approx U^0 + U^1 e^{i\delta t} + U^{-1} e^{-i\delta t} + U^2 e^{i2\delta t} + U^{-2} e^{-i2\delta t} + ... + U^K e^{iK\delta t} + U^{-K} e^{-iK\delta t}, \tag{77}$$

Here, we have made use of the following matrices:

$$U^0 = \left[H^0 \tilde{\rho}^0 - \tilde{\rho}^0 \left(H^0\right)^\dagger\right] + \left(H^1 \tilde{\rho}^1 - \tilde{\rho}^1 H^1\right) + \left(H^{-1} \tilde{\rho}^{-1} - \tilde{\rho}^{-1} H^{-1}\right), \tag{78}$$

$$U^k = \left(H^1 \tilde{\rho}^{k-1} - \tilde{\rho}^{k-1} H^1\right) + \left[H^0 \tilde{\rho}^k - \tilde{\rho}^k \left(H^0\right)^\dagger\right] + \left(H^{-1} \tilde{\rho}^{k+1} - \tilde{\rho}^{k+1} H^{-1}\right), \tag{79}$$

$$U^{-k} = \left(H^{-1} \tilde{\rho}^{-k+1} - \tilde{\rho}^{-k+1} H^{-1}\right) + \left[H^0 \tilde{\rho}^{-k} - \tilde{\rho}^{-k} \left(H^0\right)^\dagger\right] + \left(H^1 \tilde{\rho}^{-k-1} - \tilde{\rho}^{-k-1} H^1\right). \tag{80}$$

$$U^K = \left(H^1 \tilde{\rho}^{K-1} - \tilde{\rho}^{K-1} H^1\right) + \left[H^0 \tilde{\rho}^K - \tilde{\rho}^K \left(H^0\right)^\dagger\right], \tag{81}$$

$$U^{-K} = \left(H^{-1} \tilde{\rho}^{-K+1} - \tilde{\rho}^{-K+1} H^{-1}\right) + \left[H^0 \tilde{\rho}^{-K} - \tilde{\rho}^{-K} \left(H^0\right)^\dagger\right], \tag{82}$$

where $k = 1, 2, 3, ..., K-1$.

In equivalence to Eqs. (25)-(27) we can rewrite Eq. (76) in three equations as:

$$G^0 \equiv -\frac{i}{\hbar}U^0 + \tilde{\rho}_s^0 = 0, \tag{83}$$

$$G^k \equiv -\frac{i}{\hbar}U^k + \tilde{\rho}_s^k - ik\delta\tilde{\rho}^k = 0, \tag{84}$$

$$G^{-k} \equiv -\frac{i}{\hbar}U^{-k} + \tilde{\rho}_s^{-k} + ik\delta\tilde{\rho}^{-k} = 0. \tag{85}$$

As mentioned in Section 2, there are only three unique matrices when evaluating the pseudo-commutator while setting the density matrix element with the same subscripts to unity. For example, when setting $\tilde{\rho}_{ij}^0$, $\tilde{\rho}_{ij}^1$, $\tilde{\rho}_{ij}^{-1}$, …, $\tilde{\rho}_{ij}^K$, or $\tilde{\rho}_{ij}^{-K}$ to unity, we make use of the $\Lambda$ matrix as defined in Eq. (57). The three unique matrices for the pseudo-commutator are in the same form as Eq. (58) to Eq. (60). With these three matrices, all of the values of $U$ can be determined when setting $\tilde{\rho}_{ij}^0$, $\tilde{\rho}_{ij}^1$, $\tilde{\rho}_{ij}^{-1}$, …, $\tilde{\rho}_{ij}^K$, or $\tilde{\rho}_{ij}^{-K}$ to unity. By evaluating these three matrices, the efficiency of the calculation can be improved dramatically, especially when $K$ is a large number.

To obtain the pseudo-steady state solution of the density matrix, we again construct the $A$ vector from the density matrix and solve Eq. (8). The ordering for the elements in the $A$ vector follows the same pattern as that described in Section 2, which yields:

$$A \equiv \left[\tilde{\rho}_{11}^0, \tilde{\rho}_{11}^1, \tilde{\rho}_{11}^{-1}, \tilde{\rho}_{11}^2, \tilde{\rho}_{11}^{-2}, ..., \tilde{\rho}_{11}^K, \tilde{\rho}_{11}^{-K}, \tilde{\rho}_{12}^0, \tilde{\rho}_{12}^1, \tilde{\rho}_{12}^{-1}, ..., \tilde{\rho}_{NN}^K, \tilde{\rho}_{NN}^{-K}\right]^{\mathrm{T}}. \tag{86}$$

The dimension of the $A$ vector is $(2K+1)N^2 \times 1$. The ordering of the linear equations follows the same pattern. The process for evaluating the $M$ matrix, which has a dimension of $(2K+1)N^2 \times (2K+1)N^2$, is the same as that presented in Section 2. Specifically, we set an element in the $A$ vector to unity and the other elements to zeroes.

The rest of the numerical approach remains the same. Specifically, we apply the closed system constraints and acquire the $M'$ matrix and the $B'$ vector. By solving Eq. (12), we can determine the $A'$ vector and recover the $A$ vector, which represents the pseudo steady state solution for the density matrix for a system with arbitrary number of energy levels while keeping up to an arbitrary order or time oscillating terms.

# Appendix B: MATLAB codes for the numerical approach

```
clear;
%% Parameters
kB = 1.38e-23;                              % Boltzmann constant
h = 6.626e-34;                              % Planck's constant
c0 = 3e8;                                   % Speed of light
IsatD1 = 66.76;                             % Saturated intensity in W/(m^2)
IsatD2 = 43.283;
Lambda0 = 795e-9;
w0 = c0 * 2 * pi / (Lambda0);
kD1 = 2 * pi / (Lambda0);
n0 = 3e18;
Temp = 273.15 + 100;
amu2kg = 1.66e-27;                          % Factor between amu and kg
mRb = 84.912;                               % Rb-85 atomic mass

%% Decay rates
GammaD1 = 2*pi*1e7;                         % Decay rate of D1 in rad/s
GammaOP = 2*pi*0e7;
AsusD1 = h/ (2*pi) * c0 * n0 * 0.5*GammaD1 / IsatD1;     % Used for calculating susceptibility

%% Resolution control
NDelS = 501;                                % Number of sampling points
NVA = 0e3+1;                                % Number of velocity groups for Doppler effect
N = 2;                                      % Number of levels
N_order = 3;                                % Number of highest order terms kept

%% Initialize matrices
Ham = zeros(N, N);                          % Hamiltonian
density = zeros(N, N);                      % Density matrix
M = zeros((1+N_order*2)*N^2, (1+N_order*2)*N^2);         % M matrix
Q = zeros(N, N);                            % Q matrix
A = zeros((1+N_order*2)*N^2, NDelS);                     % A vector
B = zeros((N^2-(1+N_order*2)), NDelS);                   % B vector
S = zeros((N^2-(1+N_order*2)), 1);                       % S vector
W = zeros(((1+N_order*2)*N^2-(1+N_order*2)), ((1+N_order*2)*N^2-(1+N_order*2))); % W matrix

chiS = zeros(1, NDelS);
TotSusS = zeros(1,NDelS);
chiP = zeros(1, NDelS);
TotSusP = zeros(1,NDelS);

%% Doppler broadening/velocity averaging
if NVA == 1
    Velocity = 0;
    VWeight = 1;
else
    VSigma = (2*kB*Temp/(mRb*amu2kg))^0.5;
    VMin = - 5*VSigma;
    VMax = + 5*VSigma;
    Velocity = linspace(VMin,Vmax,NVA);
```

```
    VDis = exp(-Velocity.^2./VSigma.^2);
    VNor = sum(VDis);
    VWeight = 1./VNor.*VDis;
end
DeltaVD1 = kD1.*Velocity;

%% Rabi frequencies
OmegaS = 2*pi*6e6;
OmegaP = 2*pi*36e6;

%% Define detuning
DeltaP0 = 2*pi*0e7;
DelSMin = - 2*pi*1.5e8;
DelSMax = 2*pi*1.5e8;
DeltaS = linspace(DelSMin,DelSMax,NDelS);
DeltaS_norm = (DeltaS) ./(2*pi) ;                       % Transfer rad into Hz

%% Hamiltonian
%  H = [0,                       OmegaP/2*Exp0 + OmegaS/2*ExpP;
%      OmegaP/2*Exp0 + OmegaS/2*ExpM,    (-DeltaS-DeltaVD1-0.5i*GammaD1)*Exp0]
%  Exp0 = 1
%  ExpP = exp(i*Delta*t)
%  ExpM = exp(-i*Delta*t)

% ExpP=1,Exp0=ExpM=0
HP = [0,         OmegaS/2; ...
      0,         0];
% ExpM=1,Exp0=ExpP=0
HM = [0,         0; ...
      OmegaS/2,  0];

%% Main loop
for nDelS = 1 : NDelS
  Delta = DeltaS(1,nDelS);
  for nVA = 1 : NVA
    % Exp0=1,ExpP=ExpM=0
    H0 = [-0.5i*GammaOP,   OmegaP/2; ...
          OmegaP/2,        -DeltaP0-DeltaVD1(1,nVA)-0.5i*GammaD1];

    %% Solve Liouville equation
    for p = 1 : N^2
      % Find indices for Q
      remainder = rem(p, N);
      if remainder == 0
         beta = N;
      else
         beta = remainder;
      end
      alpha = ( 1 + (p-beta) / N);
      for q = 1 : N^2
         remainder = rem(q, N);
```

```matlab
            if remainder == 0
                sigma = N;
            else
                sigma = remainder;
            end
            eps = (1 + (q-sigma) / N);
            % Set a certain term to 1
            density = zeros(N,N);
            density(eps,sigma) = 1;
            % Source matrix
%             Psource = [GammaD1*density(2,2),   0; ...
%                        0,                   GammaOP*density(1,1)];
            % Seperate different factors
            Q0 = (-1i).*(H0*density-density*conj(H0));
            QP = (-1i).*(HP*density-density*conj(HP));
            QM = (-1i).*(HM*density-density*conj(HM));
            % Exp0
            M((1+N_order*2)*(p-1)+1,(1+N_order*2)*(q-1)+1)=Q0(alpha,beta);
            
            for n_order = 1:N_order
                M((1+N_order*2)*(p-1)+1,(1+N_order*2)*(q-1)+(n_order-1)*N_order+2)=QM(alpha,beta);
                M((1+N_order*2)*(p-1)+1,(1+N_order*2)*(q-1)+(n_order-1)*N_order+3)=QP(alpha,beta);
            end
            
            if (eps == alpha) && (sigma == beta)
                for n_order = 1:N_order
                    % ExpP
                    M((1+N_order*2)*(p-1)+(n_order-1)*2+2,(1+N_order*2)*(q-1)+(n_order-1)*2+2)=Q0(alpha,beta)-(1i*Delta*n_order);
                    % ExpM
                    M((1+N_order*2)*(p-1)+(n_order-1)*2+3,(1+N_order*2)*(q-1)+(n_order-1)*2+3)=Q0(alpha,beta)+(1i*Delta*n_order);
                end
            else
                for n_order = 1:N_order
                    % ExpP
                    M((1+N_order*2)*(p-1)+(n_order-1)*2+2,(1+N_order*2)*(q-1)+(n_order-1)*2+2)=Q0(alpha,beta);
                    % ExpM
                    M((1+N_order*2)*(p-1)+(n_order-1)*2+3,(1+N_order*2)*(q-1)+(n_order-1)*2+3)=Q0(alpha,beta);
                end
            end
            
            % Zeroth order to first order
            % ExpP
            M((1+N_order*2)*(p-1)+2,(1+N_order*2)*(q-1)+1)=QP(alpha,beta);
            M((1+N_order*2)*(p-1)+1,(1+N_order*2)*(q-1)+2)=QM(alpha,beta);
            % ExpM
            M((1+N_order*2)*(p-1)+3,(1+N_order*2)*(q-1)+1)=QM(alpha,beta);
            M((1+N_order*2)*(p-1)+1,(1+N_order*2)*(q-1)+3)=QP(alpha,beta);
```

```matlab
            % Cross products between higher orders
            if N_order>1
                for n_order = 1:(N_order-1)
                    M((1+N_order*2)*(p-1)+(n_order-1)*2+2,(1+N_order*2)*(q-1)+n_order*2+2)=QM(alpha,beta);
                    M((1+N_order*2)*(p-1)+(n_order-1)*2+3,(1+N_order*2)*(q-1)+n_order*2+3)=QP(alpha,beta);

                    M((1+N_order*2)*(p-1)+n_order*2+2,(1+N_order*2)*(q-1)+(n_order-1)*2+2)=QP(alpha,beta);
                    M((1+N_order*2)*(p-1)+n_order*2+3,(1+N_order*2)*(q-1)+(n_order-1)*2+3)=QM(alpha,beta);
                end
            end

            %% Source terms
            if (eps==2) && (sigma==2) && (alpha==1) && (beta==1)
                for n_order = 1:(N_order*2+1)
                    M((1+N_order*2)*(p-1)+n_order,(1+N_order*2)*(q-1)+n_order)=M((1+N_order*2)*(p-1)+n_order,(1+N_order*2)*(q-1)+n_order)+GammaD1;
                end
            end
            if (eps==1) && (sigma==1) && (alpha==2) && (beta==2)
                for n_order = 1:(N_order*2+1)
                    M((1+N_order*2)*(p-1)+n_order,(1+N_order*2)*(q-1)+n_order)=M((1+N_order*2)*(p-1)+n_order,(1+N_order*2)*(q-1)+n_order)+GammaOP;
                end
            end
          end
        end

        S = -M(1:((1+N_order*2)*N^2-(1+N_order*2)), (1+N_order*2)*N^2-(1+N_order*2)+1);
        W = M(1:((1+N_order*2)*N^2-(1+N_order*2)), 1:((1+N_order*2)*N^2-(1+N_order*2)));
        W(:,1:(1+N_order*2))=W(:,1:(1+N_order*2))-M(1:((1+N_order*2)*N^2-(1+N_order*2)),((1+N_order*2)*N^2-(1+N_order*2)+1):(1+N_order*2)*N^2);
        B = W\S;
        A(1:length(B),nDelS) = B;
        A((1+N_order*2)*N^2-(1+N_order*2)+1,nDelS) = 1-B(1);
        for n_order = 2:(1+N_order*2)
            A((1+N_order*2)*N^2-(1+N_order*2)+n_order,nDelS) = -B(n_order);
        end
        chiS(1,nDelS) = -AsusD1 * (0.5*GammaD1 / OmegaS)* A((2-1)*N*(1+2*N_order)+3,nDelS);
        TotSusS(1,nDelS) = TotSusS(1,nDelS) + VWeight(1,nVA) * chiS(1,nDelS);
    end
end

%% Plot
figure(1);
set(gcf,'position',[900,200,700,600]);
plot(DeltaS_norm.*1e-6,-imag(TotSusS), DeltaS_norm.*1e-6,real(TotSusS));
```

```
set(gca,'FontSize',16);
xlabel('Detuning (MHz)','FontName','Times New Roman','FontSize',20);
ylabel('-{\chi}'''',{\chi}'' (probe)','FontName','Times New Roman','FontSize',20);
legend('-{\chi}''''','{\chi}''');
grid on;
```

# Appendix C: MATLAB codes for the symbolic approach

```matlab
%% System parameters
Gamma = 2*pi*10^7;      % Decay rate of the excited state 1/s
Omega1 = 2*pi*36*10^6;  % Rabi frequency of the pump 1/s
Omega2 = 2*pi*6*10^6;   % Rabi frequency of the probe 1/s
Delta = 0;              % Pump detuning 1/s
rop = 0;                % Pumping rate from ground state to the excited state 1/s
N=2;                    % Number of the energy levels
Num=501;                % Number of detuning points

Temp = 273.15+100;
kB = 1.38064852 * 10^(-23);
hbar =1.0545718*10^(-34);
Natom = 3*10^18      % Atomic density 1/m^3
co = 3*10^8;         % Speed of light m/s
Isat = 120;          % Saturation intensity W/m^3

%% Symbols declare
syms  Y Z delta
rho0 = sym('rho%d0%d0', [N N]);
rho1 = sym('rho%d0%d1', [N N]);
rhom1 = sym('rho%d0%dm1', [N N]);
eqall = sym('eqall%d',[1 3*N^2]);

rho = rho0+rho1*Y+rhom1*Z;

%% Defining the Hamiltonian and source term

H = [-1i*rop/2 , -1/2*(Omega1+Omega2*Y);
   -1/2*(Omega1+Omega2*Z) , -Delta-1i*Gamma/2];

rhoSource = [rho(2,2)*Gamma,0 ; 0,rho(1,1)*rop];

Hdagger  = H-diag(diag(H))+diag(conj(diag(H)));

%% The right hand side of Liouville equation
rhodot = -1i*(H*rho-rho*Hdagger)+rhoSource;

rhodotVec = reshape(rhodot.',[],1);
paramsPlus = reshape(rho1.',[],1);
paramsMinus= reshape(rhom1.',[],1);
paramsNot= reshape(rho0.',[],1);
rhoExp =[paramsNot,paramsPlus,paramsMinus];

%% Equations separation according to the exponent order
 for j=1:N^2
   eqm = rhodotVec(j)-(1i*delta*paramsPlus(j)*Y-1i*delta*paramsMinus(j)*Z);
   [c,t] = coeffs(eqm,[Y,Z]);
```

```matlab
        indexY = find(t==Y);
        indexZ = find(t==Z);
        indexProd = find(t==Y*Z);
        index1 = find(t==1);

    if isempty (indexProd)
    eqall(j*3-2)=c(index1)==0;
    else
     eqall(j*3-2)=c(indexProd)+c(index1)==0;
    end
     eqall(j*3-1)=c(indexY)==0;
     eqall(j*3)=c(indexZ)==0;
  end

  rhoall = reshape(rhoExp.',[],1);
  %%% Ordering the set of linear equations in a matrix form
  [M,B] = equationsToMatrix(eqall,rhoall);

  %%% Defining the reduced matrix M' and organize the equations accordingly
   S     =      [M(1:(3*N^2-3),3*N^2-2:3*N^2-2),M(1:(3*N^2-3),3*N^2-1:3*N^2-1),M(1:(3*N^2-3),3*N^2:3*N^2)];
   W = M(1:(3*N^2-3),1:(3*N^2-3));

      for d=1:N-1
         W(:,((d-1)*N+d)*3-2) = W(:,((d-1)*N+d)*3-2) -S(:,1);
         W(:,((d-1)*N+d)*3-1) = W(:,((d-1)*N+d)*3-1)-S(:,2);
         W(:,((d-1)*N+d)*3) = W(:,((d-1)*N+d)*3)-S(:,3);
      end

 SolVec= B(1:3*N^2-3,1)-M(1:(3*N^2-3),3*N^2-2:3*N^2-2);
 sol = (W\SolVec);     % The reduced rho vector as a function of the symbol delta

delta = linspace(-1*150*10^6,1*150*10^6,Num)*2*pi;
result = subs(sol);    % Evaluate the sol for each delta in the input array

rho211Sol =result(N*3+3,:);  % The density matrix element associated with probe susceptibility
chi =hbar*co*Natom/(Isat*Omega2)*(Gamma/2)^2*rho211Sol ;

%%% Plot figure
figure(1)
plot(delta/2/pi/10^6,-imag(chi));
hold on
plot(delta/2/pi/10^6,real(chi));
grid on
xlabel('Detuning (MHz)','FontName','Times New Roman','FontSize',20);
ylabel('-{\chi}''',{\chi}" (probe)','FontName','Times New Roman','FontSize',20);
legend('-{\chi}"','{\chi}''');
text(-130,3.5*10^-4,'(d)','FontName','Times New Roman','FontSize',20)
```